\documentclass[journal,twoside]{IEEEtran}
\pdfoutput=1
\usepackage{mathrsfs}

\usepackage[noadjust]{cite}
\usepackage{graphicx,color,overpic,psfrag}
\usepackage{amsmath, amssymb}
\usepackage{latexsym}
\usepackage{bm}
\usepackage{amssymb}
\usepackage{cases}
\usepackage{array}
\usepackage{fancyhdr}
\usepackage{setspace}

\ifCLASSOPTIONcompsoc
\usepackage[caption=false,font=normalsize,labelfont=sf,textfont=sf]{subfig}
\else
\usepackage[caption=false,font=footnotesize]{subfig}
\fi

\usepackage{url}
\usepackage{algpseudocode}
\usepackage{algorithm}
\usepackage{blkarray}
\usepackage{booktabs}

\usepackage{multirow}
\usepackage{dsfont}
\usepackage{tabularx}
\usepackage[table]{xcolor}

\usepackage{amsfonts}

\usepackage{letltxmacro}

\graphicspath{{figure/}}

\newcommand{\ra}[1]{\renewcommand{\arraystretch}{#1}}

\newtheorem{prop}{Proposition}

\newcommand{\figref}[1]{Fig. \ref{#1}}
\newcommand{\tabref}[1]{Table \ref{#1}}

\newcommand{\appref}[1]{Appendix \ref{#1}}

\newcommand{\propref}[1]{Proposition \ref{#1}}

\newcommand{\thetabs}[2]{{\dnnot{\theta}{bs}}}

\newcommand{\bd}{\mathbf{d}}

\newcommand{\bp}{\mathbf{p}}
\newcommand{\bq}{\mathbf{q}}

\newcommand{\bY}{\mathbf{Y}}

\newcommand{\bPhi}{{\boldsymbol\Phi}}

\newcommand{\dnnot}[2]{#1_{\mathrm{#2}}}

\newcommand{\ntb}{\notag\\}

\newcommand{\figdoucolwid}{0.5\textwidth}

\allowdisplaybreaks

\begin{document}
	\title{Location-Based Timing Advance Estimation for 5G Integrated LEO Satellite Communications}

\author{
Wenjin~Wang,~\IEEEmembership{Member,~IEEE},
Tingting~Chen,
Rui~Ding,
Gonzalo~Seco-Granados,~\IEEEmembership{Senior Member,~IEEE},
Li~You,~\IEEEmembership{Member,~IEEE}, and Xiqi~Gao,~\IEEEmembership{Fellow,~IEEE}%
\thanks{Wenjin Wang, Tingting Chen, Li You, and Xiqi Gao are with the National Mobile Communications Research Laboratory, Southeast University, Nanjing 210096, China, and also with the Purple Mountain Laboratories, Nanjing 211100, China (e-mail: wangwj@seu.edu.cn; ttchen@seu.edu.cn; liyou@seu.edu.cn; xqgao@seu.edu.cn).}
\thanks{Rui Ding is with the Institute of Telecommunication Satellite, China Academy of Space Technology, Beijing 100094, China (e-mail: greatdn@qq.com).}
\thanks{Gonzalo Seco-Granados is with the Telecommunications and Systems Engineering Department, Universitat Aut\`{o}noma de Barcelona, Barcelona 08193, Spain (e-mail: Gonzalo.Seco@uab.cat).}
}

	\maketitle
	
	\begin{abstract}
		Integrated satellite-terrestrial communications networks aim to exploit both the satellite and the ground mobile communications, thus providing genuine ubiquitous coverage. 
		For 5G integrated low earth orbit (LEO) satellite communication systems, the timing advance (TA) is required to be estimated in the initial random access procedure in order to facilitate the uplink frame alignment among different users.
		However, due to the inherent characteristics of LEO satellite communication systems, e.g., wide beam coverage and long propagation delays, the existing 5G terrestrial uplink TA scheme is not applicable in the satellite networks. 
		In this paper, we investigate location-based TA estimation for 5G integrated LEO satellite communication systems. We obtain the time difference of arrival (TDOA) and frequency difference of arrival (FDOA) measurements in the downlink timing and frequency synchronization phase, which are made from the satellite at different time instants. We propose to take these measurements for either UE geolocation or ephemeris estimation, thus calculating the TA value. The estimation is then formulated as a quadratic optimization problem whose globally optimal solution can be obtained by a quadratic penalty algorithm.
		To reduce the computational complexity, we further propose an alternative approximation method based on iteratively performing a linearization procedure on the quadratic equality constraints.
		Numerical results show that the proposed methods can approach the constrained Cram\'{e}r-Rao lower bound (CRLB) of the TA estimation and thus assure uplink frame alignment for different users.
	\end{abstract}

	%
	\section{Introduction}
	
	\IEEEPARstart{D}{uring} the recent years, with the standardization of the 5G new radio (NR) communication systems and the ongoing resurgence of satellite communications (SatCom), the integration of satellite and terrestrial 5G networks is considered as a promising approach for future mobile communications   \cite{Integ,Satellite-enabled,Energy,Satellite-5G,kato19optim,Jia18space}. 
	Thanks to the wide-area service coverage capabilities, satellite networks are expected to foster the roll out 5G services in un-served areas that are not covered by terrestrial 5G networks \cite{3gpp.38.811,ASurvey,robust18wang,outa19you,massive20you,sat18kap}. 
	Several key impacts in 5G NR protocols/architecture have been identified to provide support for non-terrestrial networks \cite{Architectures,ran18xiong,Kons18use}, one of which is the adaptability of the existing 5G uplinking timing advance (TA) method in low earth orbit (LEO) SatCom.  
	  
	To ensure the uplink intra-cell orthogonality, 5G NR requires that the signals transmitted from different users within the same subframe arrive approximately in a time-aligned manner  when reaching the base station (BS), i.e., the BS can receive the uplink frames within the range of one cyclic prefix (CP) \cite{sch185g,Erik18nr,3gpp.38.213}.
	To this end, 5G NR employs an uplink TA scheme during the random access procedure to avoid timing misalignment interference, particularly in the terrestrial networks.
    However, in a typical LEO SatCom system, the differential time delay will be significantly larger than that of the terrestrial networks.\footnote{For example, with an orbital altitude of 1000 km and the minimum elevation angle of $ 20^{\circ} $, the differential time delay will be approximately 3.74 ms.} Moreover, the propagation delay in a satellite-to-ground link varies due to the fast movement of the LEO satellite. 
	Such a significant difference between the LEO SatCom system and terrestrial wireless one raises a question: \textit{Is it possible to achieve accurate TA estimation in the LEO satellite networks employing a random access procedure compatible with 5G NR?} This paper aims to answer this question.
 
	The TA estimation for random access in non-terrestrial networks (NTN) has been investigated during the past few years. 
	Recent 3rd generation partnership project (3GPP) studies have identified that location information of user equipment (UE) is beneficial for uplink TA estimation \cite{3gpp.38.811}. 
	Some proposals also consider several physical random access channel (PRACH) formats for long-distance transmissions, such as the use of long sequences (length $ = $839) for both FR1 (450 MHz-6 GHz) and FR2 (24.25 GHz-52.6 GHz) operating bands, more repetition or multiple sequence transmissions \cite{ran19harri,ran18zhen,Zhen20prea,Si13lte,Caus20new,3gpp.38.104}. 
	For SatCom systems with large Doppler shifts and oscillator uncertainties, symmetric Zadoff-Chu (ZC) sequences have been adopted to estimate TA \cite{Enhanced}. 
	In \cite{Two-step}, a two-step time delay difference estimation was presented for SatCom systems, which first divides a beam cell into some layered small sub-areas and then two types of PRACH preamble burst formats are transmitted. 
	TA estimation based on the correlation between a ZC sequence and its conjugate replica has been used in \cite{Timing}. In \cite{Yu20timing}, a reliable TA estimation approach with robustness to frequency offset was proposed in satellite mobile communication scenarios. Compared with sending TA commands from the satellite to the UE, signaling overhead can be significantly reduced if the TA value can be estimated directly at the UE side. However, to the best of our knowledge, most previous works on TA estimation for SatCom systems were carried out at the satellite side during the uplink while little focus has been placed on the investigation of TA estimation at the UE side with the utilization of 5G downlink synchronization signals.
	  
    In this paper, we propose a novel UE location information-assisted approach for uplink TA estimation in 5G integrated LEO SatCom. Two specific but important scenarios are taken into account. One is that the satellite broadcasts ephemeris periodically, and the UE is not able to obtain its position using a global navigation satellite system (GNSS). The visibility of 5G LEO satellites and GNSS satellites, which are mostly deployed in the medium earth orbits (MEOs), is not similar \cite{Prin13paul}. Moreover, sometimes, it is expected to design a system that can work independently of the other systems. Hence, the consideration of this scenario is reasonable. This could happen in urban scenarios where having enough GNSS satellites to compute the position is troublesome, but 5G LEO satellites are visible. The other is that the satellite does not broadcast ephemeris, and the UE can perform GNSS positioning. 
	We utilize the timing and frequency offset estimates acquired from downlink synchronization signals to perform TA estimation in these two scenarios separately. 
	The timing and frequency offset estimation can be then transformed into time difference of arrival (TDOA) and frequency difference of arrival (FDOA) measurements equivalently. As the combined TDOA and FDOA measurements have been extensively used in the source localization \cite{Sun11an,Kc04an,Geolocation,Iterative}, we propose to adopt them to estimate the UE location or satellite ephemeris information, and then the value of uplink TA can be calculated at the UE side. The major contributions of our work are summarized as follows:
	  \begin{itemize}
	  	\item We propose a method for TA estimation in 5G integrated LEO SatCom through a 5G NR compatible random access procedure. Depending on whether the UE has the capability of GNSS positioning or not, we divide the scenarios for TA estimation problem into two categories and carry out the studies separately. We further extend the problem into the multi-satellite networks based on the system model of TA estimation in the single-satellite case.
	  	
	  	\item We estimate the timing and frequency offset in the downlink synchronization phase to acquire the TDOA and FDOA measurements. With these measurements, we convert the problem of TA estimation into either UE geolocation or ephemeris estimation. As the altitude of UE is often known, we exploit this altitude information to improve the positioning accuracy of UE.
	  	
	  	\item We formulate the equality-constrained optimization problem via using the system model of 5G integrated LEO SatCom for either geographical location or ephemeris estimation. Then we propose a quadratic penalty algorithm to find the globally optimal solution of the estimation problem. In order to reduce the computational complexity, we further propose an iterative constrained weighted least squares (CWLS) method for this equality-constrained problem.
	  	
	  \end{itemize}
	  
	  Some of the notations adopted in this paper are listed as follows:
	  \begin{itemize}
	  	
	  \item Upper and lower case boldface letters denote matrices and column vectors, respectively. 
	  \item $ \mathbb{R}^{M\times N} $ denotes the $ M \times N $ dimensional real-valued vector space.
	  \item $\mathbf{I}_{N}$ and $ \mathbf{0}_{M\times N} $ denote the $ N\times N $ dimensional identity matrix and $ M\times N $ dimensional zero matrix, respectively. The subscripts are sometimes omitted for brevity.
	  \item The superscript $ (\cdot)^{T} $, $ (\cdot)^{-1} $, and $ (\cdot)^{\dagger} $ denote the transpose, inverse, and pseudo-inverse operations, respectively. 
	  \item $ \lVert \cdot \lVert $ denotes the Euclidean norm. 
	  \item $ \mathrm{det}{(\cdot)} $ denotes the determinant operation. 
	  \item $\dot {(\cdot)}$ denotes the derivative of $(\cdot)$ with respect to time.
	  \item $ \mathrm{diag}(\cdot) $ denotes the diagonal matrix with the elements of $ (\cdot) $ on the main diagonal. 
	  \item $\nabla(\cdot) $ denotes the gradient computation. 
	  \item $ [\mathbf{A}]_{i,:} $, $ [\mathbf{A}]_{:,j} $, and $ [\mathbf{A}]_{i,j} $ denote the $ i $-th row, the $ j $-th column and the $ (i,j) $-th element of the matrix $ \mathbf{A} $, respectively.
	  \item All estimated parameters are described as $ \hat{(\cdot)} $.
	  \end{itemize}
	  
	  The rest of the paper is organized as follows. Section \ref{system_m} introduces the TA problem in 5G integrated LEO SatCom. Section \ref{location_esti} and Section \ref{ephemeris_esti} present the location and ephemeris estimation algorithms with downlink synchronization signals, respectively. Section \ref{multi-satellite} presents the TA estimation in multi-satellite systems. Section \ref{numerical_resu} illustrates the numerical results and Section \ref{conclusion} concludes the paper.
	
	\section{System Model}\label{system_m}
      \subsection{Timing Advance in 5G Integrated LEO SatCom}
      
      In the development of 5G integrated SatCom, most existing works focused on the air interface design of the satellite module to maximize utilization of the technology commonalities with the terrestrial systems, so as to reduce the implementation costs and simplify the interactive procedures.
      For example, as the 5G NR basic waveform, CP Orthogonal Frequency Division Multiplexing Access (CP-OFDMA) requires that the signals transmitted from different UEs are time-aligned when reaching the BS to keep the uplink intra-cell orthogonality, i.e., any timing misalignment across the received signals should fall within the range of one CP.
      To this end, 5G NR adopts a scheme for TA during the random access procedure, where the BS first estimates the uplink TA for the UE through the PRACH preamble and then sends the adjustment information to the UE by the random access response (RAR) message \cite{3gpp.38.213}. 
      The UE further adjusts its uplink transmission time based on the received TA values combined with the acquired downlink timing synchronization information. 
      
      However, such a scheme is designed specifically for the terrestrial networks and may not be applicable to the satellite-to-ground transmission.
      For example, the cell coverage is limited by the CP length of PRACH preambles. 
      The current 5G NR PRACH preambles with $L_{RA} = 839$ formats allow the cell coverage varying from 15 km to 102 km. The beam footprints of LEO satellites can be designed to be similar in size to the terrestrial cell coverage. However, in most practical LEO satellites, due to the consideration of coverage requirements and implementation complexity, the diameter of beam footprints is designed on the order of several hundreds of kilometers, which is significantly larger than that in terrestrial systems \cite{Por19tech,Xia19beam}.
      In addition, large variation of the round-trip delay in satellite-to-ground communications within a cell/beam would limit the availability of cyclic shift (CS) multiplexing as well, resulting in smaller cell reuse factor of preamble sequences \cite{R1.1908818}.
      On the other hand, the maximum value of the TA command in the RAR message defined in 5G NR may be smaller than the round-trip delay of a LEO satellite-to-ground link \cite{3gpp.38.811}.   
      \newcolumntype{L}{>{\hspace*{-\tabcolsep}}l}
      \newcolumntype{R}{c<{\hspace*{-\tabcolsep}}}
      \definecolor{lightblue}{rgb}{0.93,0.95,1.0}
 
\begin{figure}
      \centering
      \includegraphics[width=8cm]{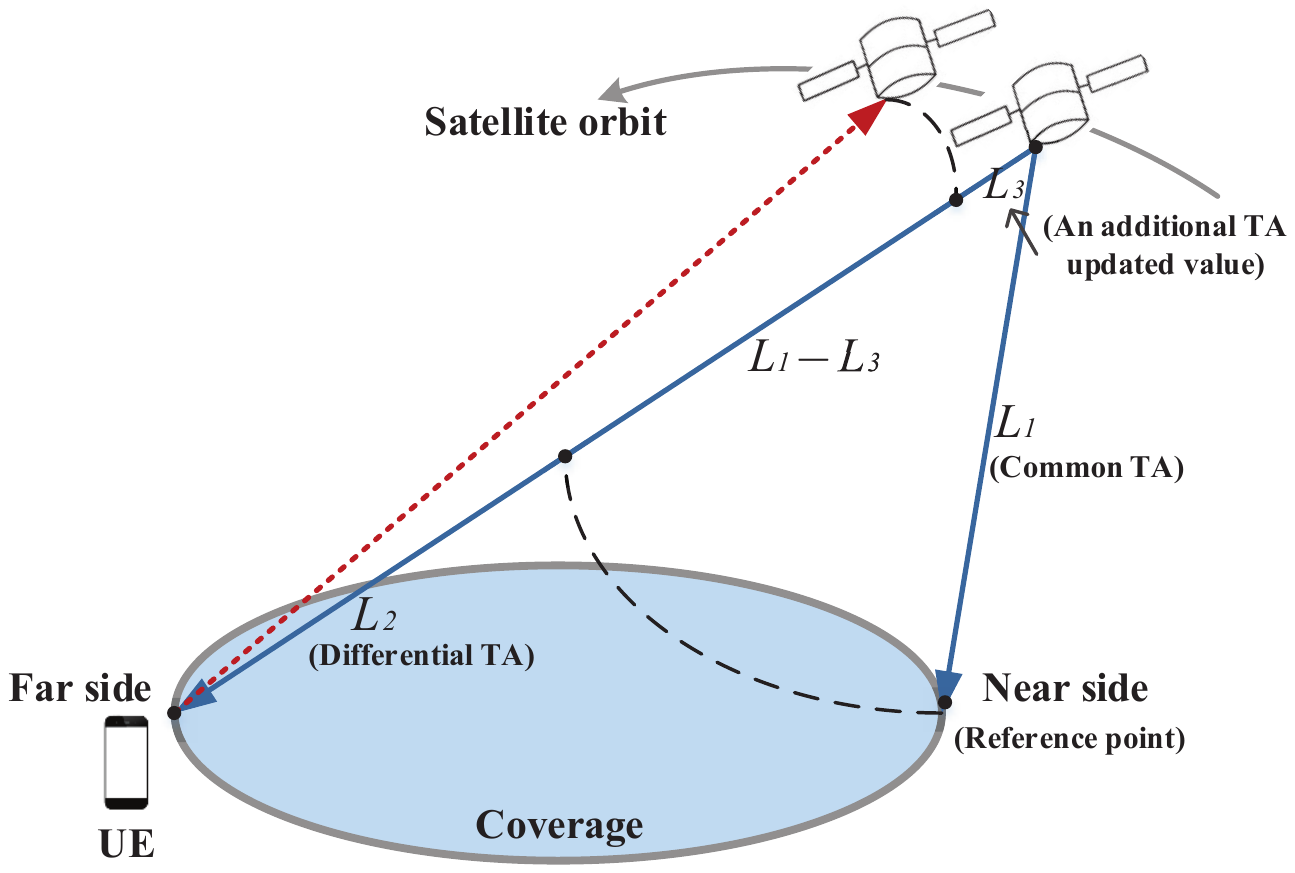}
      \caption{Illustration of cell/beam coverage of NTN.}
      \label{fig1_0325}
      \end{figure}

      In addition, due to the high-speed motion of the LEO satellite, the satellite-to-ground links usually exhibit a varying propagation delay. As this change rate of the propagation delay between the UE and satellite is very large, the TA command sent by the satellite is outdated by the time the UE receives it \cite{R1.1910982}. Hence, to account for this expected TA inaccuracy, an additional adjustment procedure is needed to update the original TA command, as shown in Fig. \ref{fig1_0325}. 
      
      In view of these challenges, 3GPP has agreed that several options can be considered to support TA adjustment in random access procedure for NTN. 
      Firstly, when UE positioning capabilities, e.g., GNSS positioning, are enabled at the UE side, they can be used to enhance the TA estimation at the UE side and minimize the amount of signaling required, especially in LEO SatCom systems \cite{R1.1910064}. 
      Moreover, as indicated by Fig. \ref{fig1_0325}, the total TA can be divided into beam/cell specific common TA $ L_{1} $ and user-specific differential TA $ L_{2} $, where the former is used to compensate for the round-trip delay at a reference point within the cell/beam, e.g., the nearest point to the satellite, and the latter is used to represent the difference between the common TA and the actual TA for a specific user. 
      Note that the common TA can be obtained by UEs via broadcast information from the satellite and then only the differential TA is responded in the RAR message.
      
      \subsection{Location Based Timing Advance Estimation}
      
      In this part, we investigate how a UE can get a rough user-specific TA in 5G integrated LEO SatCom using only downlink signals. In 5G NR, the downlink synchronization signal block (SSB) is introduced, consisting of the primary and secondary synchronization signals \cite{3gpp.38.300}. The UE can acquire timing and frequency synchronization within a cell and the physical layer Cell ID through the detection of SSB. In the following, we first introduce the 5G downlink synchronization signals received at the UE side.
      
      Consider $ K $ OFDM symbols in one frame. Denote the number of subcarriers and length of CP as $ N $ and $ N_{g} $, respectively. Then the received signal corresponding to the $ k $-th OFDM symbol $ s_{k}(n),k=1,2,...,K,n=0,1,...,N+N_{g}-1 $ can be given by \cite{Near}
      \begin{align}\label{eq1}
      	r_{k}(n)=e^{j2\pi n\varepsilon/N}\sum_{l=0}^{L-1}h(l)s_{k}(n-\theta-l)+z(n),
      \end{align}
      where $\theta$ denotes the integer-valued symbol timing offset (normalized by the sampling interval) and $\varepsilon$ denotes the normalized carrier frequency offset (CFO) with respect to the subcarrier spacing, including the contribution of the difference in the transmitter and receiver oscillators $ \varepsilon_{e} $ and the Doppler frequency shift $ \varepsilon_{d} $ introduced during the downlink transmission of signals. $ \varepsilon_{e} $ is assumed to be constant during the observation time of the signal \cite{Morelli16robust,Hsieh99low,Minn03robust}. In addition, $ h(l) $, $ l=0,1,...,L-1 $ denotes the impulse response of a multipath channel with $ L $ uncorrelated taps, and $ z(n) $ is the additive white Gaussian noise with zero mean and variance $ \sigma_{z}^{2} $.
       
      By using the maximum log-likelihood criterion, discrete prolate spheroidal sequences (DPSS), and the CP structure of OFDM, the timing and frequency offset estimation algorithms in \cite{Near} can achieve near optimal performance, i.e., accurate estimation of parameters $ \theta $ and $ \varepsilon $ based on the observation $ r_{k}(n) $ at the receiver is available. Denote the total number of downlink SSBs as $ M $. Consider a short period, namely the timing-window,\footnote{Note that the timing-window refers to the product of ($ M-1 $) and the time interval between two adjacent SSBs.} in which the satellite is visible to the target ground UE. Within this timing-window, we can then obtain sequences of timing and  frequency offset estimates corresponding to different SSBs written as $ 
      \widetilde{\theta}_{i} $ and $ \widetilde{\varepsilon}_{i} $, where $ i=1,2,...,M $ denotes the serial number of SSBs.
      Consider the initial synchronization timing offset $ \widetilde\theta_{1} $ and CFO $ \widetilde\varepsilon_{1} $ as the reference, then the estimated TDOA $ \widetilde t_{i,1} $ and FDOA $\widetilde f_{i,1} $, $ i=2,3,...,M $ between SSB $ i $ and SSB $ 1 $ are given by
      \begin{subequations}\label{equ1}
      \begin{align}
      \widetilde t_{i,1}&=(\widetilde\theta_{i}-\widetilde \theta_{1}) T_{s},\label{equ1_1}\\
      \widetilde f_{i,1}&=(\widetilde \varepsilon_{i}-\widetilde \varepsilon_{1}) \Delta f,\label{equ1_2}
      \end{align}
      \end{subequations}
      respectively, where $ T_{s} $ and $ \Delta f $ represent the sampling interval and subcarrier spacing, respectively. Note that TDOA noise $ \Delta t_{i,1} $ and FDOA noise $ \Delta f_{i,1} $ caused by estimation error of the timing offset and CFO can be written as
      \begin{subequations}\label{ti1}
      \begin{align}
      \widetilde t_{i,1}=t_{i,1}+\Delta t_{i,1},\\
      \widetilde f_{i,1}=f_{i,1}+\Delta f_{i,1},
      \end{align}
      \end{subequations}
      where $ t_{i,1} $ and $ f_{i,1} $ denote the noise-free values of TDOA and FDOA, respectively.
      
      Hence, based on the timing and CFO estimation algorithm with 5G downlink synchronization signals, we can obtain noisy measurements of TDOA and FDOA.
      Then, we perform the problem of TA estimation in the following two important scenarios:
      
      \begin{itemize}
      	\item[1)]\textit{Scenario 1: The satellite broadcasts ephemeris periodically and the GNSS service is not available for UEs.} As the ephemeris information is available at the UE side, the TA estimation for random access for the 5G integrated LEO SatCom is therefore transformed into the location estimation of the UE with the utilization of downlink synchronization signals. With the UE location estimates and ephemeris information, the propagation delay between UE and the satellite can be estimated at the UE side. The UE can then adjust the timing of its uplink transmissions based on the delay estimates.
      	Regarding the broadcasting ephemeris data, there are two different possible representations as described in \cite{3gpp.38.821}. One possibility is to use orbital parameters, including the orbital plane and satellite level parameters. The other is to provide the location and velocity of the satellite combined with a reference point in time. Since several satellites typically share a common orbital plane in a satellite network, orbital plane parameters remain the same and can be pre-provisioned to UE as baseline ephemeris data. Then only satellite level parameters are required to be broadcasted to UE via system information. Hence, in this paper, we consider the first option to represent the ephemeris data to reduce the broadcasting overhead. In addition, to make sure that the UE always uses the latest ephemeris data for initial access, once the UE has obtained new ephemeris data, the parameters stored in the UE are obsolete and overwritten by the newer values.
      	
      	\item[2)]\textit{Scenario 2: The satellite does not broadcast ephemeris but the UE has the capability of GNSS positioning.} In this scenario, UE location is available through the GNSS positioning. However, the ephemeris information is unknown. Thus, the TA estimation is converted into the ephemeris estimation with the downlink synchronization signals, and then the propagation delay between UE and the satellite can be calculated. 
      \end{itemize}
 
\begin{figure}
      	\centering
      	\includegraphics[width=9cm]{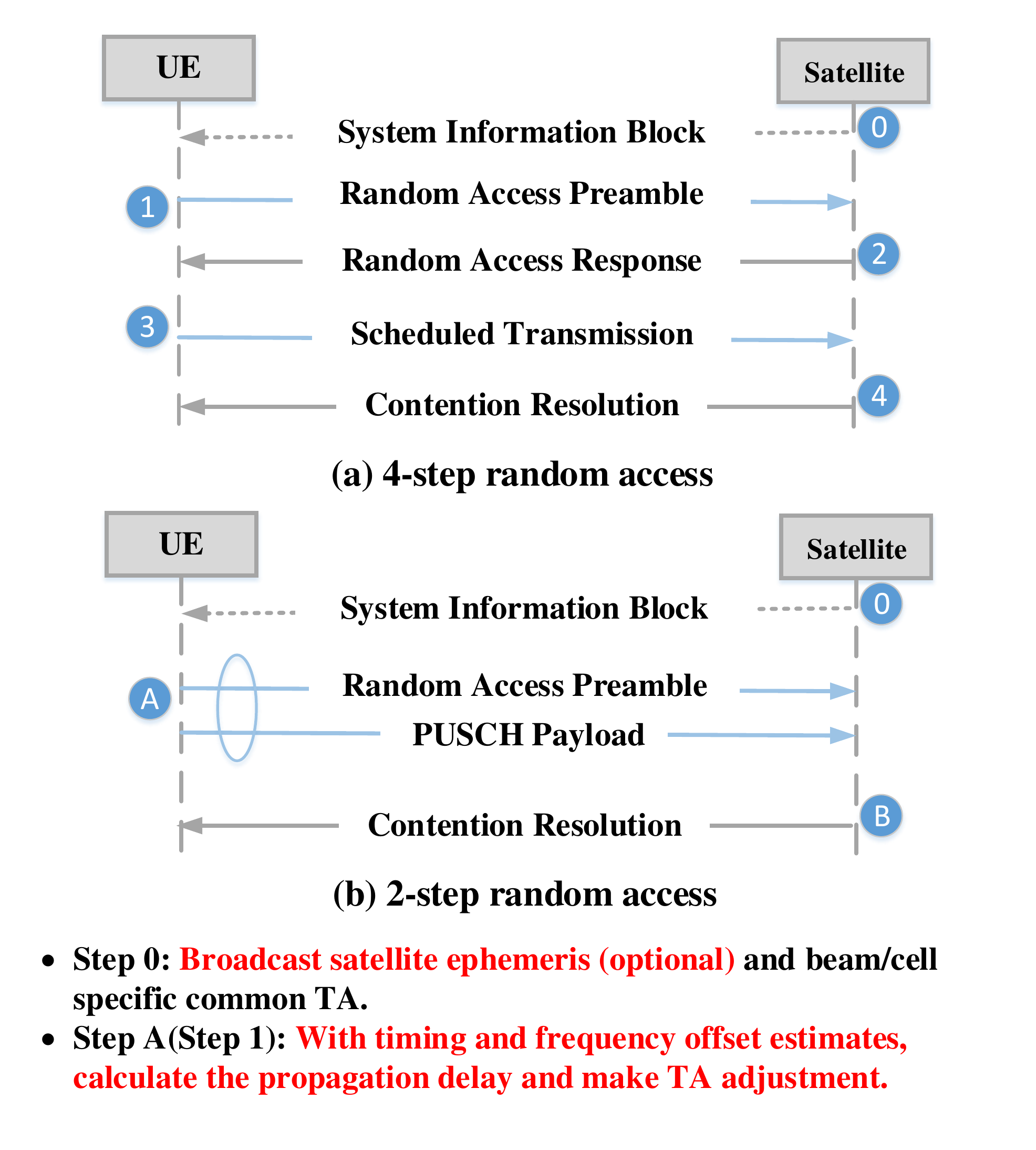}
      	\caption{The random access procedure with location-based TA estimation.}
      	\label{fig1_0203}
      \end{figure}

      Note that in this paper, we perform the TA estimation method in the regenerative architecture, which only considers the user link between the satellite and UE. In practice, the gateway and satellite locations are assumed to be known by each other. Then the propagation delay and frequency shift of the feeder link can be pre-compensated at the satellite or the gateway side. Hence, our work can also be applied to the transparent architecture. In addition, both the 2-step and 4-step random access procedures are taken into consideration in this paper. Fig. \ref{fig1_0203} shows the 2-step and 4-step random access procedures with location-based TA estimation for 5G integrated LEO SatCom. The type of random access is selected based on the network configuration by the UE at initiation of the random access procedure.
      As the TA estimation is transformed into the geolocation of UE and the ephemeris estimation with downlink synchronization signals, next sections further give the location estimation algorithms for \textit{Scenario 1} and ephemeris estimation algorithms for \textit{Scenario 2}.
	\section{Location Estimation Algorithms with Downlink Synchronization Signals}\label{location_esti}
	
	In this section, we focus on \textit{Scenario 1}, and our aim is to estimate the UE location with the downlink synchronization signals. With the relationship between timing/frequency offset estimation and TDOA/FDOA measurements in (\ref{equ1_1}) and (\ref{equ1_2}), UE location estimation with the downlink synchronization signals is converted into the geolocation with joint TDOA and FDOA measurements. In this part, we first relate the TDOA and FDOA measurements to the unknown UE location.
	
	\subsection{Problem Formulation}
	
    \begin{figure}
    	\centering
    	\includegraphics[width=8cm]{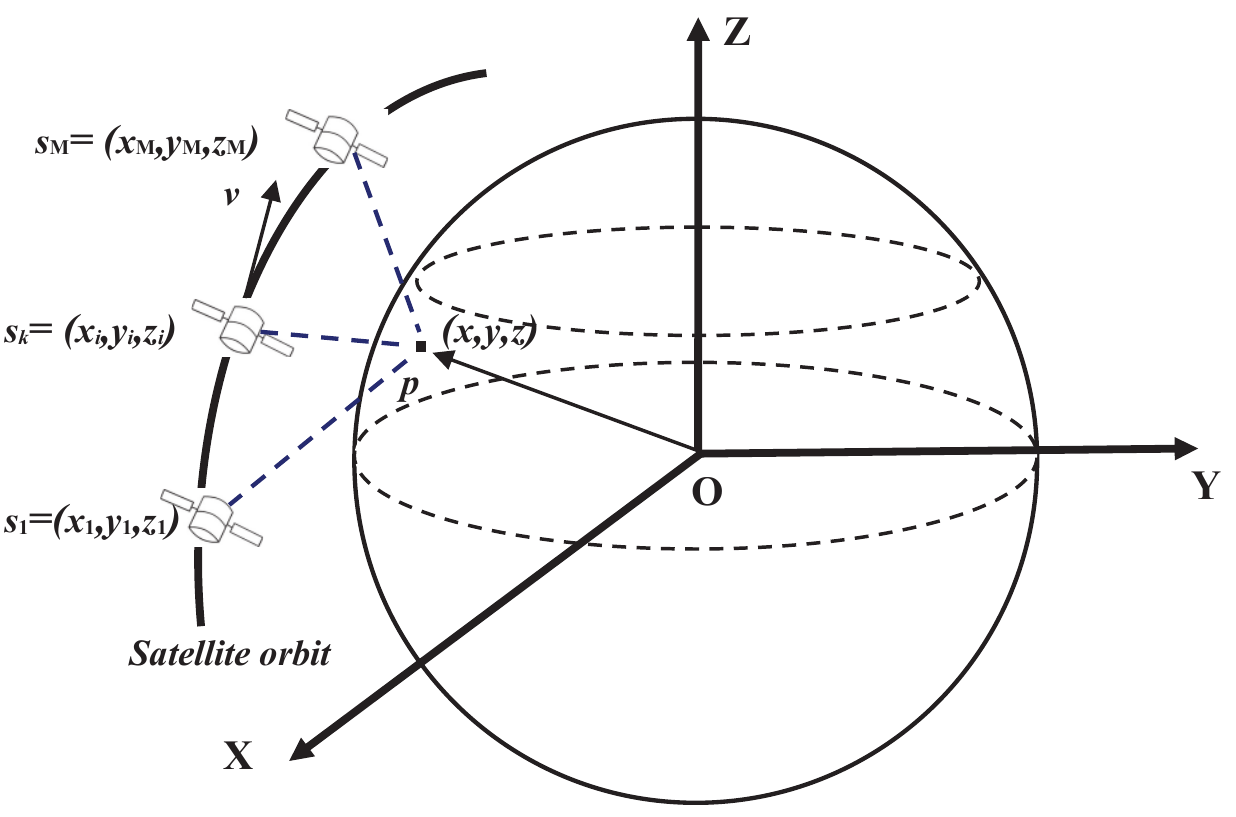}
    	\caption{Geometric relation of a single LEO satellite and the UE in ECEF coordinate.}
    	\label{relpos} 
    \end{figure}
    
	Fig. \ref{relpos} shows a single LEO satellite with a UE located on the surface of the earth in the Earth Centered Earth Fixed (ECEF) coordinate system, which is aligned with the equatorial plane and the Greenwich meridian \cite{SatelliteOrbits}. The position and velocity vector of the UE in ECEF are denoted by $\mathbf{p}=[x,y,z]^{T}$ and $\dot{\mathbf{p}}=[\dot{x},\dot{y},\dot{z}]^{T}$, respectively. The satellite locations $\mathbf{s}_{i}=[x_{i},y_{i},z_{i}]^{T}$ and velocities $ \dot{\mathbf{s}}_{i}=[\dot{x}_{i},\dot{y}_{i},\dot{z}_{i}]^{T} $ when at the transmit instant of the $ i $-th SSB, $ i=1,2,...,M $ are assumed to be known thanks to the broadcast satellite ephemeris as supposed in \textit{Scenario 1}.
	
	Let $d_{i}$ represent the distance between the satellite and the UE corresponding to the $ i $-th downlink SSB given by
	\begin{align}\label{di}
	d_{i}=\rVert\mathbf{s}_{i}-\mathbf{p}\rVert,\quad i=1,2,...,M.
	\end{align}
	Then the range difference of arrival between the $ i $-th and the first SSB related to the TDOAs is given by 
	\begin{align}\label{di1}
	d_{i,1}=d_{i}-d_{1}=c t_{i,1},\quad i=2,3,...,M,
	\end{align}
	where $ c $ is the speed of light. Taking derivative of (\ref{di1}) with respect to time, we can obtain the range rate differences \cite{Iterative}, which are denoted as $ \dot{d}_{i,1} $ given by
	\begin{align}\label{di11}
	\dot d_{i,1}=c \dot t_{i,1}=\dot d_{i}-\dot d_{1},\quad i=2,3,...,M,
	\end{align}
	where $ \dot t_{i,1} $ and $ \dot d_{i} $ denote the rate of change of $ t_{i,1} $ and $ d_{i} $, respectively. From the derivative of (\ref{di}) with respect to time, $ \dot d_{i} $ can be further described as
	\begin{align}\label{ddi}
	\dot d_{i}
	=\frac{(\mathbf{s}_{i}-\mathbf{p})^{T}(\dot{\mathbf{s}}_{i}-\dot{\bp})}{d_{i}}.
	\end{align}
	$ \dot t_{i,1} $ in (\ref{di11}) can be derived from the FDOAs written as \cite{pc82emitter}
	\begin{align}\label{fi1}
	f_{i,1}=f_{c}{\dot t_{i,1}},
	\end{align}
	where $ f_{c} $ denotes the carrier frequency. 
	
	From (\ref{di1}), it can be observed that TDOAs are equivalent to the range differences. In the following, TDOAs and the range differences will be used interchangeably. In addition, FDOAs and range rate differences will be also used interchangeably as they are equivalent by (\ref{di11}) and (\ref{fi1}).
	
	Taking into account the influence of noises caused by estimation errors of timing and frequency offsets, we define $ \widetilde{d}_{i,1} $ and $ \widetilde{\dot{d}}_{i,1} $ as the measured value of range and range rate differences, respectively. They can be derived from noisy measurements of TDOA and FDOA sequences as
	\begin{subequations}
	\begin{align}
	\widetilde{d}_{i,1}&=d_{i,1}+c\Delta t_{i,1},\\ \widetilde{\dot{d}}_{i,1}&=\dot{d}_{i,1}+ c\Delta \dot{t}_{i,1}, 
	\end{align}
	\end{subequations}
	where $ \Delta \dot{t}_{i,1} = \Delta f_{i,1}/f_{c} $ is equivalent to the FDOA noise. Let $ \mathbf{n}_{t}=[c\Delta t_{2,1},c\Delta t_{3,1},...,c\Delta t_{M,1}]^{T}\in\mathbb{R}^{(M-1)\times1} $ and $ \mathbf{n}_{f}=[c\Delta \dot t_{2,1},c\Delta\dot t_{3,1},...,c\Delta \dot t_{M,1}]^{T}\in\mathbb{R}^{(M-1)\times1} $ be the vectors of TDOA and FDOA noises, respectively. We assume that they are both zero mean and have covariance matrix as follows
	\begin{align}
	\mathbf{Q}_{t}=\mathbf{E}[\mathbf{n}_{t}\mathbf{n}_{t}^{T}],\quad
	\mathbf{Q}_{f}=\mathbf{E}[\mathbf{n}_{f}\mathbf{n}_{f}^{T}].
	\end{align} 
	
	Developing the squared term in $d_{i}^{2}=(d_{i,1}+d_{1})^{2}$, and using (\ref{di}) to span $d_{i}^{2}$ and $d_{1}^{2}$, we can then obtain a set of TDOA equations
	\begin{equation}
	\begin{aligned}
	\label{ri11}
	d_{i,1}^{2} +2d_{i,1}d_{1}=-2(\mathbf{s}_{i}-\mathbf{s}_{1})^{T}\mathbf{p}+\mathbf{s}_{i}^{T}\mathbf{s}_{i}-\mathbf{s}_{1}^{T}\mathbf{s}_{1},\\ i=2,3,...,M.
	\end{aligned}
	\end{equation}
	Further, to make use of FDOAs, we take derivative of (\ref{ri11}) with respect to time and obtain
	\begin{equation}\label{td}
	\begin{aligned}
	{ d_{i, 1} \dot{d}_{i, 1}+ d_{i, 1} \dot{d}_{1}+ \dot{d}_{i, 1} d_{1}- \mathbf{s}_{i}^{T} \dot{\mathbf{s}}_{i}+ \mathbf{s}_{1}^{T} \dot{\mathbf{s}}_{1}}  =\\-\left(\dot{\mathbf{s}}_{i}-\dot{\mathbf{s}}_{1}\right)^{T} \mathbf{p}-\left(\mathbf{s}_{i}-\mathbf{s}_{1}\right)^{T} \dot{\mathbf{p}}.
	\end{aligned}
	\end{equation}
	
	Define $ \mathbf{u}_{1}=[\mathbf{p}^{T},\dot{\mathbf{p}}^{T},d_{1},\dot{d}_{1}]^{T} $. Note that $ d_{i,1}=\widetilde{d}_{i,1}-c\Delta t_{i,1} $, $ \dot{d}_{i,1}=\widetilde{\dot{d}}_{i,1}-c\Delta \dot{t}_{i,1} $, then the set of equations (\ref{ri11}) and (\ref{td}) becomes 
		\begin{align}\label{h1}
	\mathbf{h}_{1}=\mathbf{G}\mathbf{u}_{1}+\boldsymbol{\epsilon},
	\end{align}
	where 
	\begin{equation}
	\mathbf{h}_{1} =\left[\begin{array}{c}{\widetilde d_{2,1}^{2}-\mathbf{s}_{2}^{T} \mathbf{s}_{2}+\mathbf{s}_{1}^{T} \mathbf{s}_{1}} \\ {\widetilde d_{3,1}^{2}-\mathbf{s}_{3}^{T} \mathbf{s}_{3}+\mathbf{s}_{1}^{T} \mathbf{s}_{1}} \\ {\vdots} \\ {\widetilde d_{M,1}^{2}-\mathbf{s}_{M}^{T} \mathbf{s}_{M}+\mathbf{s}_{1}^{T} \mathbf{s}_{1}} \\ {2 \widetilde d_{2, 1} \widetilde{\dot{d}}_{2, 1}-2 \mathbf{s}_{2}^{T} \dot{\mathbf{s}}_{2}+2 \mathbf{s}_{1}^{T} \dot{\mathbf{s}}_{1}} \\ {2\widetilde d_{3, 1} \widetilde{\dot{d}}_{3, 1}-2 \mathbf{s}_{3}^{T} \dot{\mathbf{s}}_{3}+2 \mathbf{s}_{1}^{T} \dot{\mathbf{s}}_{1}} \\ {\vdots} \\ {2\widetilde d_{M, 1} \widetilde{\dot{d}}_{M, 1}-2 \mathbf{s}_{M}^{T} \dot{\mathbf{s}}_{M}+2 \mathbf{s}_{1}^{T} \dot{\mathbf{s}}_{1}}\end{array}\right],
	\end{equation}
	\begin{equation}
	\begin{aligned}
	\mathbf{G} = -2\begin{bmatrix}
	\mathbf{s}_{2}^{T}-\mathbf{s}_{1}^{T} & \mathbf{0}_{1\times3} & \widetilde{d}_{2,1} & 0 \\
	\mathbf{s}_{3}^{T}-\mathbf{s}_{1}^{T} & \mathbf{0}_{1\times3}& \widetilde{d}_{3,1} & 0 \\
	\vdots                  & \vdots & \vdots              & \vdots \\
	\mathbf{s}_{M}^{T}-\mathbf{s}_{1}^{T} & \mathbf{0}_{1\times3}& \widetilde{d}_{M,1} & 0 \\
	\dot{\mathbf{s}}_{2}^{T}-\dot{\mathbf{s}}_{1}^{T} & \mathbf{s}_{2}^{T}-\mathbf{s}_{1}^{T} & \widetilde{\dot{d}}_{2,1} &\widetilde{d}_{2,1}\\  
	\dot{\mathbf{s}}_{3}^{T}-\dot{\mathbf{s}}_{1}^{T} & \mathbf{s}_{3}^{T}-\mathbf{s}_{1}^{T}& \widetilde{\dot{d}}_{3,1} &\widetilde{d}_{3,1}\\
	\vdots            &\vdots      & \vdots              & \vdots \\
	\dot{\mathbf{s}}_{M}^{T}-\dot{\mathbf{s}}_{1}^{T} & \mathbf{s}_{M}^{T}-\mathbf{s}_{1}^{T}& \widetilde{\dot{d}}_{M,1} &\widetilde{d}_{M,1}
	\end{bmatrix},
	\end{aligned}
	\end{equation}
	and $ \boldsymbol{\epsilon} $ is the error vector derived from (\ref{ri11}) and (\ref{td}). By ignoring the second order error term, $ \boldsymbol{\epsilon} $ becomes a Gaussian random vector with covariance matrix given by
	\begin{align}\label{psi}
	\boldsymbol\Psi=\left[\begin{array}{cc}{\mathbf{B}} & {\mathbf{0}} \\ {\dot{\mathbf{B}}} & {\mathbf{B}}\end{array}\right]\left[\begin{array}{cc}{\mathbf{Q}_{t}} & {\mathbf{0}} \\ {\mathbf{0}} & {\mathbf{Q}_{f}}\end{array}\right]\left[\begin{array}{cc}{\mathbf{B}} & {\dot{\mathbf{B}}} \\ {\mathbf{0}} & {\mathbf{B}}\end{array}\right]\\ \nonumber
	\in\mathbb{R}^{2(M-1)\times2(M-1)},
	\end{align}
	where
	\begin{align}\label{Bdiag}
	\mathbf{B}=&2\mathrm{diag}\{d_{2},d_{3},...,d_{M}\}\in\mathbb{R}^{(M-1)\times(M-1)},\\\label{dBdiag}
	\dot{\mathbf{B}}=&2\mathrm{diag}\{\dot{d}_{2},\dot{d}_{3},...,\dot{d}_{M}\}\in\mathbb{R}^{(M-1)\times(M-1)}.
	\end{align} 
	
	Consider that the elements of $ \mathbf{u}_{1} $ are statistically independent, then the maximum-likelihood estimation of $ \mathbf{u}_{1} $ can be written as
	\begin{align}\label{esti_1}
	\mathbf{\hat u}_{1}=\mathop{\arg\max}\limits_{\mathbf{u}_{1}}\ \log f(\mathbf{h}_{1}|\mathbf{u}_{1}),
	\end{align}
	where $ f(\mathbf{h}_{1}|\mathbf{u}_{1}) $ is the conditional probability density function of $ \mathbf{h}_{1} $ given $ \mathbf{u}_{1} $ and given by
	\begin{align}\label{esti_2}
	f(\mathbf{h}_{1}|\mathbf{u}_{1})=&\dfrac{1}{(2\pi)^{M-1}(\mathrm{det}(\boldsymbol\Psi))^{1/2}}\ntb&\cdot\exp{\left\lbrace -\dfrac{1}{2}(\mathbf{h}_{1}-\mathbf{G}\mathbf{u}_{1})^{T}\boldsymbol\Psi^{-1}(\mathbf{h}_{1}-\mathbf{G}\mathbf{u}_{1}) \right\rbrace }.
	\end{align}
	Thus, the maximum-likelihood estimation of $ \mathbf{u}_{1} $ can be described as
	\begin{align}\label{esti}
	\mathbf{\hat u}_{1}=\mathop{\arg\min}\limits_{\mathbf{u}_{1}}\ {\left\lbrace (\mathbf{h}_{1}-\mathbf{G}\mathbf{u}_{1})^{T}\boldsymbol\Psi^{-1}(\mathbf{h}_{1}-\mathbf{G}\mathbf{u}_{1})\right\rbrace}.
	\end{align} 
	Weighting matrix $ \boldsymbol\Psi $ is unknown in practice as $\mathbf{B}$ and $ \dot{\mathbf{B}} $ contain the accurate satellite-UE distance and its rate of change, respectively.
	We propose to solve this problem through a further approximation, which considers two typical cases. In the first case of short-time random access procedure, $d_{i} (i=1,...,M)$ is close to each other. Supposing they all approach $d^{0}$, then $\mathbf{B}\approx 2d^{0}\mathbf{I}$ is satisfied. Correspondingly, $\dot{\mathbf{B}}\approx \mathbf{0}$ is also satisfied. Since scaling $ \boldsymbol{\Psi} $ does not affect the solution to problem (\ref{esti}), we substitute $ \mathbf{I} $ for $\mathbf{B}$ to simplify the weighting matrix. In the other case of initial access for the search of the UE location, since the observation window of the satellite might be much larger, we take the initial value of $\mathbf{B}$ and $ \dot{\mathbf{B}} $ as $ \mathbf{I} $ and $ \mathbf{0} $, respectively, and then iteratively update the weighting matrix by (\ref{psi}) with the latest estimation results \cite{Geolocation}. In the following, we focus on the estimation problem with a fixed weighting matrix. 
	
	In the solution of problem (\ref{esti}), the correlations among the elements of $ \mathbf{u}_{1} $ are not considered. However, they are related to each other in practice \cite{Feng16app,Feng18on}. In the following, we aim to exploit this relationship to provide an improved estimate. Firstly, consider a non-spherical earth model \cite{Geolocation} and that the UE is located on the surface of the earth, then the UE location $ \mathbf{p} $ satisfies the following equation
	\begin{align}\label{earth}
	\frac{{x}^{2}}{(R_{a})^{2}}+\frac{{y}^{2}}{(R_{a})^{2}}+\frac{{z}^{2}}{(R_{b})^{2}}-1=0,
	\end{align}
	where $R_{a}$ and $ R_{b} $ denote the semi-major and semi-minor axes of the earth, respectively. In addition, the elements of $ \mathbf{u}_{1} $ are also related by (\ref{di}) and (\ref{ddi}) at $ i=1 $. 
	With these constraints, the above TDOA/FDOA based location estimation problem in (\ref{esti}) can be reformulated as
	\begin{align}\label{obj1}
	\mathop{\mathrm{minimize}}\limits_{\mathbf{u}_{2}}\quad & g(\mathbf{u}_{2})=(\mathbf{h}_{2}-\mathbf{G}\mathbf{u}_{2})^{T}\boldsymbol \Psi^{-1}(\mathbf{h}_{2}-\mathbf{G}\mathbf{u}_{2}), \nonumber\\
	\mathrm{subject}\ \mathrm{to} \quad& c_{1}(\mathbf{u}_{2})=\mathbf{u}_{2}^{T}\mathbf{C}_{1}\mathbf{u}_{2}+2\mathbf{q}_{1}^{T}\mathbf{u}_{2}-\rho_{1}=0, \nonumber\\
	&c_{2}(\mathbf{u}_{2})=\mathbf{u}_{2}^{T}\mathbf{C}_{2}\mathbf{u}_{2}=0,\nonumber\\
	&c_{3}(\mathbf{u}_{2})= \mathbf{u}_{2}^{T}\mathbf{C}_{3}\mathbf{u}_{2}=0,
	\end{align} 
	where 
	\begin{subequations}
		\begin{align}
		\mathbf{u}_{2}&=\mathbf{u}_{1}-\widetilde{\mathbf{r}}_{1},\\
		\widetilde{\mathbf{r}}_{1}&=(\mathbf{s}_{1}^{T},\dot{\mathbf{s}}_{1}^{T},0,0)^{T},\\
		\mathbf{h}_{2}&=\mathbf{h}_{1}-\mathbf{G}\widetilde{\mathbf{r}}_{1},\\
		\mathbf{q}_{1}&=\mathbf{C}_{1}\widetilde{\mathbf{r}}_{1},\\
		\rho_{1}&=1-\widetilde{\mathbf{r}}_{1}^{T}\mathbf{C}_{1}\widetilde{\mathbf{r}}_{1},\\
		\mathbf{r}&=[\frac{1}{R_{a}^{2}},\frac{1}{R_{a}^{2}},\frac{1}{R_{b}^{2}}]^{T},\\
		\mathbf{C}_{1}&=\mathrm{diag}{[\mathbf{r}^{T},0,0,0,0,0]},\\
		\mathbf{C}_{2}&=\mathrm{diag}{[1,1,1,0,0,0,-1,0]},\\
		\mathbf{C}_{3}&=\left(\begin{array}{cccc}{\mathbf{0}_{3 \times 3}} & {\mathbf{I}_{3}} & {\mathbf{0}_{3 \times 1}} & {\mathbf{0}_{3 \times 1}} \\ {\mathbf{0}_{3 \times 3}} & {\mathbf{0}_{3 \times 3}} & {\mathbf{0}_{3 \times 1}} & {\mathbf{0}_{3 \times 1}} \\ {\mathbf{0}_{1 \times 3}} & {\mathbf{0}_{1 \times 3}} & {0} & {-1} \\ {\mathbf{0}_{1 \times 3}} & {\mathbf{0}_{1 \times 3}} & {0} & {0}\end{array}\right).
		\end{align}
	\end{subequations}
	
	\subsection{Globally Optimal Solution}
	
	The optimization problem in (\ref{obj1}) is a quadraric programming with quadratic equality constraints. Quadratic penalty method is commonly used in practice to solve the equality-constrained problems because of its simplicity \cite{Jor04Num}. The idea of the quadratic penalty method is to transform the original constrained problem into an equivalent unconstrained one, so that the algorithms, e.g., Newton's method and conjugate gradient method can be used to solve the equivalent unconstrained problem  \cite{nonli06Ba,nonli99Di}. In this part, we first adopt the quadratic penalty method to solve this problem and then discuss its optimality. 

    The penalty function can be defined as follows
    \begin{align}\label{Fza}
    F(\mathbf{u}_{2};\mu)=g(\mathbf{u}_{2})+\mu\alpha(\mathbf{u}_{2}),
    \end{align}
    where 
    \begin{align}\label{alp}
    \alpha(\mathbf{u}_{2})=\sum_{i=1}^{3}c_{i}^{2}(\mathbf{u}_{2}).
    \end{align}
    Note that (\ref{alp}) is the penalty function of the constraint problem (\ref{obj1}). Obviously,
    \begin{align}\label{alp1}
    \alpha(\mathbf{u}_{2})\left\{ \begin{array}{ll}
    =0,\ &\mathbf{u}_{2} \in \mathcal{D},\\
    >0,\ &\mathbf{u}_{2} \notin \mathcal{D}.
    \end{array}\right.
    \end{align}
    where $\mathcal{D}$ is the feasible set of (\ref{obj1}).
    
    Note that $F(\mathbf{u}_{2};\mu)$ is called the augmented objective function of (\ref{obj1}) where $\mu(>0)$ is the penalty parameter of the quadratic penalty method. Then, (\ref{obj1}) can be transformed into the following unconstrained optimization problem
    \begin{align}\label{uncon}
    \mathop{\mathrm{minimize}}\limits_{\mathbf{u}_{2}}\quad &F(\mathbf{u}_{2};\mu).
    \end{align} 
    
    \begin{prop}\label{equa}
    	For a given $ \boldsymbol\Psi $ and $\mu_{k}$, suppose that $\mathbf{u}_{2,k}$ is the minimum point of (\ref{uncon}). Then the sufficient and necessary condition under which $\mathbf{u}_{2,k}$ is the minimum point of (\ref{obj1}) is $\mathbf{u}_{2,k} \in \mathcal{D}$.
    \end{prop}

    \begin{IEEEproof}
    	Please refer to Appendix \ref{appendixa}.
    \end{IEEEproof}
    
    \propref{equa} shows that if the global minimizer of $F(\mathbf{u}_{2};\mu)$ belongs to the feasible set $\mathcal{D}$, then it is indeed the solution of (\ref{obj1}). 
    
    \begin{prop}\label{conv}
    	For a given weighting matrix $ \boldsymbol\Psi $, suppose that $\mathbf{u}_{2,k}$ is the global minimizer of $F(\mathbf{u}_{2};\mu_{k})$ defined by (\ref{Fza}) and that $\mu_{k}\to\infty$. Then every limit point $\mathbf{u}_{2}^{*}$ of the sequence $\{\mathbf{u}_{2,k}\}$ is a globally optimal solution to problem (\ref{obj1}).
    \end{prop}
    \begin{IEEEproof}
    	Please refer to Appendix \ref{appendixb}.
    \end{IEEEproof}
    
    To make the solution of (\ref{uncon}) approach the optimal solution to the original problem, the penalty parameter $ \mu $ is required to be sufficiently large. Therefore, we choose an increasing sequence of penalty parameters $ \{\mu_{k}\} $ to repeatedly solve a sequence of problems in (\ref{uncon}). On this basis, we adopt Newton's method \cite{Jor04Num} to solve the unconstrained optimization problem (\ref{uncon}). The iterative equation of each step is given by
    \begin{align}
    \mathbf{u}_{2,j+1}&=\mathbf{u}_{2,j}+\beta_{j}\mathbf{p}_{j},\ntb
    \mathbf{p}_{j}&=-\mathbf{G}_{j}^{-1}\nabla F_{j},	\ntb
    F_{j}&=g(\mathbf{u}_{2,j})+\mu\alpha(\mathbf{u}_{2,j}),
    \end{align}
    where $ \mathbf{u}_{2,j} $ and $ \mathbf{u}_{2,j+1} $ denote the $ j $-th and $ (j+1) $-th iteration estimated values, respectively. $\mathbf{G}_{j}$ is a nonsingular and symmetric  matrix, and the positive scalar $\beta_{j}$ is the step length for the $ j $-th iteration. To ensure the line search direction to be a descending direction, the following condition need to be satisfied: $\mathbf{p}_{j}^{T}\nabla F_{j}<0$. In Newton's method, $\mathbf{G}_{j}$ is the exact Hessian $\nabla^{2}F_{j}$. However, the Hessian matrix may not be positive definite, which causes the search direction not always to be descending. In such cases, we can adopt a modified Hessian method by adding a positive diagonal matrix to the original Hessian \cite{Jor04Num}. 
    
	\subsection{Low Complexity Method}
	
	In the above proposed algorithm, we first adopt the quadratic penalty method to transform problem  (\ref{obj1}) into an unconstrained one, and then utilize Newton's method with modification to solve the unconstrained problem in (\ref{uncon}). We show that the proposed algorithm is guaranteed to obtain a globally optimal solution. However, since the penalty parameter $\mu$ is required to be updated repeatedly and the unconstrained problem in (\ref{uncon}) for a given $ \mu $ also needs multiple iterations in Newton's method, this approach tends to have a high computational complexity. It is possible to trade-off some degree of optimality for a reduced cost. Hence, in the following, we propose an iterative CWLS method as an alternative algorithm.
	
	The method starts from an initial estimate of $ \mathbf{u}_{2} $. Then, we write one of the variables $ \mathbf{u}_{2} $ in $ c_{i}(\mathbf{u}_{2}),i=1,2,3 $ as a combination of the estimated value $ \mathbf{\hat u}_{2} $ and the estimated error $ \Delta \mathbf{u}_{2}$.
	With the estimated value of $ \mathbf{u}_{2} $, we convert the problem in (\ref{obj1}) into an approximate quadratic programming with linear equality constraints, which is verified to have a closed-form solution in \cite{Iterative}. Next, we update the linear equality constraints with the latest estimation of $ \mathbf{u}_{2} $ and solve the approximate quadratic programming iteratively. In the following, we present more detailed descriptions of this method.
	
	An initial estimate of $ \mathbf{u}_{2} $ can be calculated by (\ref{esti}) as
	\begin{align}
	\mathbf{\hat u}_{2}=(\mathbf{G}^{T}\boldsymbol\Psi^{-1}\mathbf{G})^{-1}\mathbf{G}^{T}\boldsymbol\Psi^{-1}\mathbf{h}_{2}.
	\end{align}
	Then, the approximate quadratic programming with linear equality constraints based on (\ref{obj1}) can be formulated as
	\begin{align}\label{obj111}
	\mathop{\mathrm{minimize}}\limits_{\mathbf{u}_{2}}\quad &g(\mathbf{u}_{2})=
	(\mathbf{h}_{2}-\mathbf{G}\mathbf{u}_{2})^{T}\boldsymbol \Psi^{-1}(\mathbf{h}_{2}-\mathbf{G}\mathbf{u}_{2}), \ntb
	\mathrm{subject}\ \mathrm{to} \quad& c_{1}(\mathbf{u}_{2})=(\mathbf{\hat u}_{2}^{T}\mathbf{C}_{1}+2\mathbf{q}_{1}^{T})\mathbf{u}_{2}-\rho_{1}=0, \ntb
	& c_{2}(\mathbf{u}_{2})=\mathbf{\hat u}_{2}^{T}\mathbf{C}_{2}\mathbf{u}_{2}=0,\ntb
	& c_{3}(\mathbf{u}_{2})=\mathbf{\hat u}_{2}^{T}\mathbf{C}_{3}\mathbf{u}_{2}=0.
	\end{align}
	The above problem (\ref{obj111}) has been proved to possess a closed-form solution \cite{Iterative}, which can be expressed as
	\begin{align}\label{cfs}
	\mathbf{\breve u}_{2}&=\\\nonumber
	&(\mathbf{P}_{1}\mathbf{G}^{T}\boldsymbol\Psi^{-1}\mathbf{G}\mathbf{P}_{1})^{\dagger}(\mathbf{G}^{T}\boldsymbol\Psi^{-1}\mathbf{h}_{2}-\mathbf{G}^{T}\boldsymbol\Psi^{-1}\mathbf{G}\mathbf{P}_{2})+\mathbf{P}_{2},
	\end{align}
	where 
	\begin{subequations}
		\begin{align}
		\mathbf{P}_{1}&=\mathbf{I}-\mathbf{A}^{T}(\mathbf{A}\mathbf{A}^{T})^{-1}\mathbf{A},\\
		\mathbf{P}_{2}&=\mathbf{A}^{T}(\mathbf{A}\mathbf{A}^{T})^{-1}\boldsymbol\beta_{1},\\
		\mathbf{A}&=[(\mathbf{\hat u}_{2}^{T}\mathbf{C}_{1}+2\mathbf{q}_{1}^{T});\mathbf{\hat u}_{2}^{T}\mathbf{C}_{2};\mathbf{\hat u}_{2}^{T}\mathbf{C}_{3}],\\
		\boldsymbol\beta_{1}&=[\rho_{1};0;0].
		\end{align}
	\end{subequations}
    Note that the quadratic penalty and CWLS methods both need to update the weighting matrix. Simulations show that updating the weighting matrix once is sufficient to provide an accurate result in practice. The penalty method requires inner iteration to solve the unconstrained problem (\ref{uncon}) with Newton's method and outer iteration for the updating of the penalty parameter $ \mu $. Compared with the penalty method, the CWLS method derives a closed form solution to the approximate problem (\ref{obj111}) and only requires to update the $ \mathbf{u}_{2} $ iteratively by (\ref{cfs}). Hence, the computational complexity can be significantly reduced.

	\subsection{Cram\'{e}r-Rao Lower Bound}
	
	The CRLB is the mimimum variance that an unbiased parameter estimator can attain \cite{funda93kay}. The constrained CRLB of an unbiased estimator has been given in \cite{Asimplederivation}. Let $ \mathbf{d}=[d_{2,1},d_{3,1},...,d_{M,1}]^{T} $, $ \dot{\bd}=[\dot{d}_{2,1},\dot{d}_{3,1},...,\dot{d}_{M,1}]^{T} $, and $ \mathbf{u}=[\bp^{T},\dot{\bp}^{T}]^{T} $. Combined with the system model of this paper, the constrained CRLB for $ \mathbf{u} $ can be completely calculated as
	
	\begin{align}\label{CRLB}
	\mathrm{CRLB}(\mathbf{u})=\mathbf{J}^{-1}-\mathbf{J}^{-1} \mathbf{F}\left(\mathbf{F}^{T} \mathbf{J}^{-1} \mathbf{F}\right)^{-1} \mathbf{F}^{T} \mathbf{J}^{-1},
	\end{align}
	where

	\begin{align}
	\mathbf{F}=[{x},{y},(\frac{R_{a}^{2}}{R_{b}^{2}}) {z},0,0,0]^{T},
	\end{align}
	\begin{align}
	\mathbf{J}=\left[\begin{array}{cc}{\dfrac{\partial \mathbf{d}^{T}}{\partial \mathbf{p}}} & {\dfrac{\partial \dot{\mathbf{d}}^{T}}{\partial \mathbf{p}}}\\ 
	{\dfrac{\partial \mathbf{d}^{T}}{\partial \dot{\mathbf{p}}}} & {\dfrac{\partial \dot{\mathbf{d}}^{T}}{\partial \dot{\mathbf{p}}}}
	\end{array}\right]\left[\begin{array}{cc}{\mathbf{Q}_{t}^{-1}} & \mathbf{0} \\ \mathbf{0} & {\mathbf{Q}_{f}^{-1}}\end{array}\right]\left[\begin{array}{cc}{\dfrac{\partial \mathbf{d}}{\partial \mathbf{p}^{T}}} & {\dfrac{\partial {\mathbf{d}}}{\partial \dot{\mathbf{p}}^{T}}} \\ {\dfrac{\partial \dot{\mathbf{d}}}{\partial \mathbf{p}^{T}}} & {\dfrac{\partial \dot{\mathbf{d}}}{\partial \dot{\mathbf{p}}^{T}}} \end{array}\right],
	\end{align}
	and 
	\begin{subequations}
	\begin{align}
	\dfrac{\partial \mathbf{d}}{\partial \mathbf{p}^{T}}&=-\left[\begin{array}{c}{\left(\mathbf{s}_{2}-\mathbf{p}\right)^{T} / d_{2}-\left(\mathbf{s}_{1}-\mathbf{p}\right)^{T} / d_{1}} \\ {\left(\mathbf{s}_{3}-\mathbf{p}\right)^{T} / d_{3}-\left(\mathbf{s}_{1}-\mathbf{p}\right)^{T} / d_{1}} \\ {\vdots} \\ {\left(\mathbf{s}_{M}-\mathbf{p}\right)^{T} / d_{M}-\left(\mathbf{s}_{1}-\mathbf{p}\right)^{T} / d_{1}}\end{array}\right],\\
	\dfrac{\partial \mathbf{d}}{\partial \dot{\mathbf{p}}^{T}}&=\mathbf{0}_{(M-1)\times3},\\
	\frac{\partial \dot{\mathbf{d}}}{\partial \mathbf{p}^{T}}&=\\\nonumber
	&\left(\begin{array}{c}{\frac{\left(\mathbf{s}_{2}-\mathbf{p}\right)^{T} \dot{d}_{2}}{d_{2}^{2}}-\frac{\left(\mathbf{s}_{1}-\mathbf{p}\right)^{T} \dot{d}_{1}}{d_{1}^{2}}-\frac{(\dot{\mathbf{s}}_{2}-\dot{\bp})^{T}}{d_{2}}+\frac{(\dot{\mathbf{s}}_{1}-\dot{\bp})^{T}}{d_{1}}} \\ {\frac{\left(\mathbf{s}_{3}-\mathbf{p}\right)^{T} \dot{d}_{3}}{d_{3}^{2}}-\frac{\left(\mathbf{s}_{1}-\mathbf{p}\right)^{T} \dot{d}_{1}}{d_{1}^{2}}-\frac{(\dot{\mathbf{s}}_{3}-\dot{\bp})^{T}}{d_{3}}+\frac{(\dot{\mathbf{s}}_{1}-\dot{\bp})^{T}}{d_{1}}} \\ {\vdots} \\ {\frac{\left(\mathbf{s}_{M}-\mathbf{p}\right)^{T} \dot{d}_{M}}{d_{M}^{2}}-\frac{\left(\mathbf{s}_{1}-\mathbf{p}\right)^{T} \dot{d}_{1}}{d_{1}^{2}}-\frac{(\dot{\mathbf{s}}_{M}-\dot{\bp})^{T}}{d_{M}}+\frac{(\dot{\mathbf{s}}_{1}-\dot{\bp})^{T}}{d_{1}}}\end{array}\right),\\
	\dfrac{\partial \dot{\mathbf{d}}}{\partial \dot{\mathbf{p}}^{T}}&=\frac{\partial {\mathbf{d}}}{\partial \mathbf{p}^{T}}.
	\end{align}
    \end{subequations}
	The CRLB derived from (\ref{CRLB}) will be used for the following simulations to compare with the mean-square error (MSE) of our proposed method.
	
	\section{Ephemeris Estimation Algorithms with Downlink Synchronization Signals}\label{ephemeris_esti}
	
	For \textit{Scenario 2}, we aim to perform the ephemeris estimation with the aid of downlink synchronization signals in this section. To facilitate the formulation of the following estimation problem, we first introduce two coordinates extensively used in the satellite-to-ground links.
	\subsection{Coordinates Used in the Satellite-to-Ground Links}
	
	\begin{figure}
		\centering
		\includegraphics[width=\figdoucolwid]{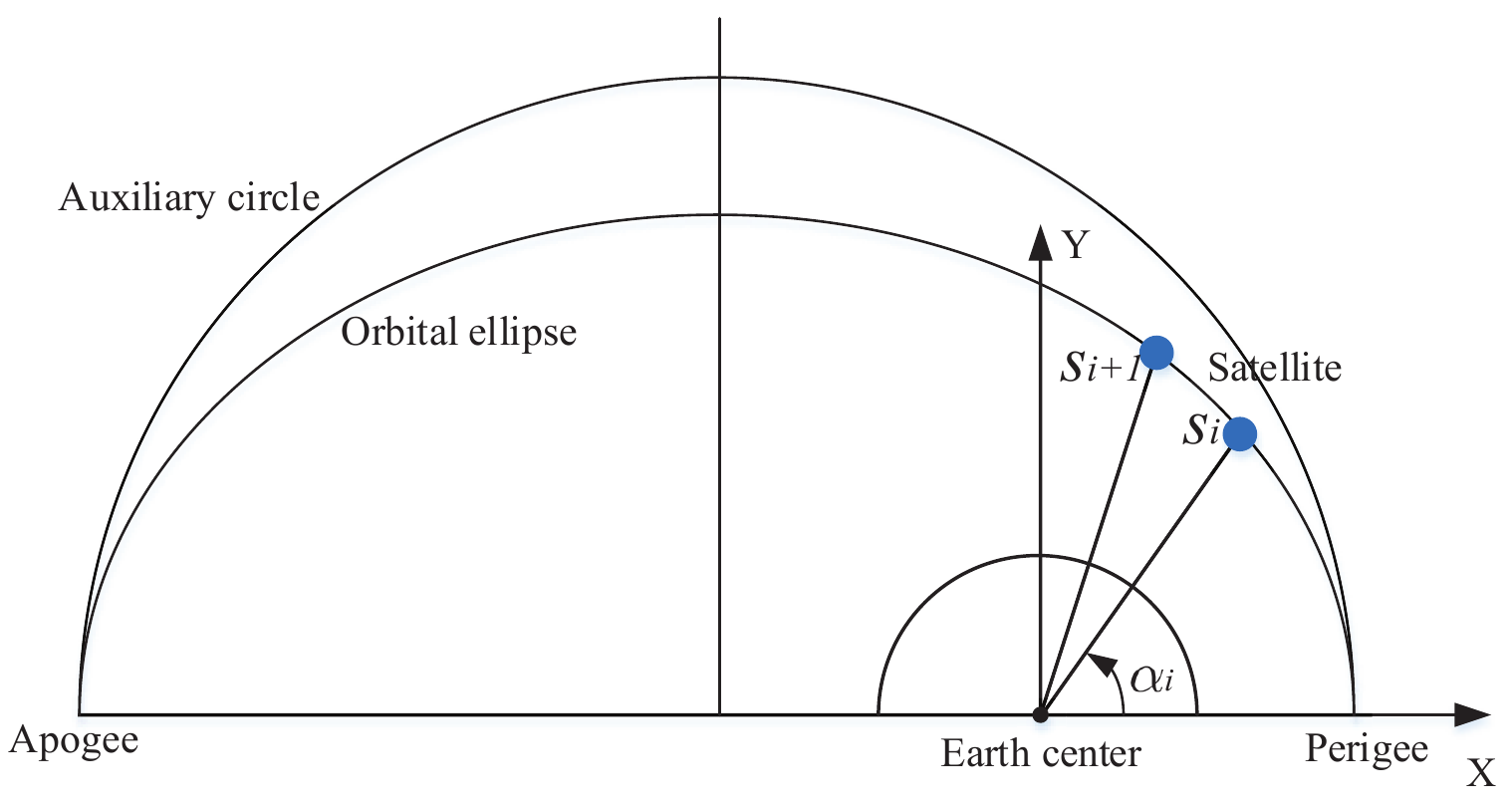}
		\caption{Orbital plane coordinate system.}
		\label{OPC} 
	\end{figure}
	
	\figref{OPC} illustrates the orbital plane coordinate (OPC), where the orbital plane forms the X-Y reference plane, X-axis points towards the perigee, and Z-axis completes a right hand rule \cite{SatelliteOrbits}. The satellite location in OPC when launching the $ i $-th SSB can be then written as
	\begin{align}\label{opc}
		(\mathbf{s}_{i})_{opc}=[r_{i}\cos\alpha_{i},r_{i}\sin\alpha_{i},0]^{T},
	\end{align}  
	where $ r_{i} $ denotes the distance between the satellite and the center of the earth, and $ \alpha_{i} $ is the angle between X-axis and position vector of the satellite as shown in \figref{OPC}. In the following, we focus on the circular orbits and denote the orbital radius by $ r $.
	
	Another common coordinate system for describing the satellite orbits is the Earth Centered Interval (ECI) coordinate with its origin at the center of the earth, X-axis aligned with the vernal equinox, and Z-axis directed to the north pole. The transformation between any two of the coordinates OPC, ECI, and ECEF can be achieved by rotations of the coordinate axes. Define the elementary matrices
	\begin{align}
	\mathbf{R_{x}}(\phi)&=\left(\begin{array}{ccc} {1} & {0} & {0}\\ {0} & {\cos\phi} & {\sin\phi}\\ {0} & {-\sin\phi} & {\cos\phi} \end{array}\right),\\
	\mathbf{R_{z}}(\phi)&=\left(\begin{array}{ccc} {\cos\phi} & {\sin\phi} & {0}\\ {-\sin\phi} & {\cos\phi} & {0}\\ {0} & {0} & {1} \end{array}\right),
	\end{align}
	to describe rotations around the X and Z axes by an angle of $ \phi $, respectively. Then, the satellite location in ECI when launching the $ i $-th SSB can be obtained by \cite{SatelliteOrbits}
	\begin{align}\label{sit}
		(\mathbf{s}_{i})_{eci}=\mathbf{R_{z}}(-\Omega)\mathbf{R_{x}}(-\vartheta)\mathbf{R_{z}}(-\varphi)(\mathbf{s}_{i})_{opc},
	\end{align}
	or
	\begin{align}\label{sit1}
	(\mathbf{s}_{i})_{eci}=\mathbf{R_{z}}(-\theta_{g_{i}})\mathbf{s}_{i},
	\end{align}
	where $ \Omega $, $ \vartheta $, and $ \varphi $ denote the right ascension of ascending node, inclination, and argument of perigee of the orbit, respectively. $ \theta_{g_{i}} $ denotes the Greenwich Sidereal Time (GST) when the satellite launches the $ i $-th SSB. 
	
	With the relationships in (\ref{sit}) and (\ref{sit1}), we next investigate how to relate the TDOAs and FDOAs to the unknown positions and velocities of the satellite.
	
	\subsection{Problem Formulation}
	
	As the GNSS service is available for UEs, the UE location and velocity are assumed known. For the notation simplicity in the following, we transform the satellite and UE coordinate from ECEF to ECI. The UE location in ECI can be described as $ (\mathbf{p}_{i})_{eci} $ given by 
	\begin{align}
		(\mathbf{p}_{i})_{eci}=\mathbf{R_{z}}(-\theta_{g_{i}})\mathbf{p}, \quad i=1,2,...,M.
	\end{align}
	Accordingly, the velocity of UE in ECI can be given by $ \dot{\mathbf{p}} $ denoted as $ (\dot{\mathbf{p}}_{i})_{eci} $.
	In the case of the definite satellite orbit, the locations of satellite corresponding to different SSBs satisfy the following equation 
	\begin{align}\label{si}
		(\mathbf{s}_{i})_{eci}=\mathbf{A}_{i}	(\mathbf{s}_{1})_{eci},\quad i=1,2,...,M,
	\end{align}
	where $ \mathbf{A}_{i} $ is a transformation matrix related to the time interval of adjacent SSBs and orbital parameters, e.g, $ \Omega $, $ \vartheta $, and $ \varphi $. The derivation of $ \mathbf{A}_{i} $ is given in \appref{appendixc}.
	
	For a satellite in a circular orbit, the velocities of the satellite in OPC at the transmit of the $ i $-th SSB can be given by
	\begin{align}\label{vopc}
	(\dot{\mathbf{s}}_{i})_{opc}=[-v\sin\alpha_{i},v\cos\alpha_{i},0]^{T},
	\end{align}  
	where $ v=\sqrt{\dfrac{\mu'}{r}} $ and $ \mu' $ is the geocentric gravitational constant. By employing the same procedure of derivation in \appref{appendixc}, the satellite velocities in ECI can be also described by the transformation matrix $ \mathbf{A}_{i} $ given by
	\begin{align}\label{dsi}
	(\dot{\mathbf{s}}_{i})_{eci}=\mathbf{A}_{i}(\dot{\mathbf{s}}_{1})_{eci}.
	\end{align}
	In addition, from (\ref{opc}) and (\ref{vopc}), the satellite velocities can be denoted by the satellite positions and the orbital parameters, i.e.,
	\begin{align}\label{sidsi}
		(\dot{\mathbf{s}}_{i})_{opc}=\mathbf{C}_{4}(\mathbf{s}_{i})_{opc},
	\end{align}
	where \begin{align}
		\mathbf{C}_{4}=\left(\begin{array}{ccc}{0} & {-v/r} & {0} \\ {v/r} & {0} & {0} \\ {0} & {0} & {0}\end{array}\right).
	\end{align}
	Equation (\ref{sidsi}) can be further rewritten as
	\begin{align}\label{dsieci}
	(\dot{\mathbf{s}}_{i})_{eci}=\bPhi(\mathbf{s}_{i})_{eci},
	\end{align}
	where 
	\begin{align}
	\bPhi&= (\mathbf{E}^{opc}_{eci})^{-1}\mathbf{C}_{4}\mathbf{E}^{opc}_{eci},\\
	\mathbf{E}^{opc}_{eci}&=\mathbf{R_{z}}(\varphi)\mathbf{R_{x}}(\vartheta)\mathbf{R_{z}}(\Omega).
	\end{align}

	With (\ref{si}), (\ref{dsi}), and (\ref{dsieci}), $ (\mathbf{s}_{i})_{eci} $ and $ (\dot{\mathbf{s}}_{i})_{eci} $ can be represented by $ (\mathbf{s}_{1})_{eci} $. Equations (\ref{ri11}) and (\ref{td}) can be then written as 
	\begin{align}\label{dii}
	d_{i,1}^{2}+2d_{i,1}d_{1}&=-2\mathbf{w}_{i}(\mathbf{s}_{1})_{eci},
	\end{align}
	and
	\begin{equation}
	\begin{aligned}\label{ddii}
	&\dot{d}_{i,1} d_{i,1}+\dot{d}_{i,1}d_{1}+d_{i,1}\dot{d}_{1}=-\mathbf{v}_{i}(\mathbf{s}_{1})_{eci},
	\end{aligned}
	\end{equation}
	where
	\begin{subequations}
	\begin{align}
	\mathbf{w}_{i}&=(\mathbf{p}_{i})_{eci}^{T}\mathbf{A}_{i}-(\mathbf{p}_{1})_{eci}^{T},\\
	\mathbf{v}_{i}&=\mathbf{w}_{i}\bPhi+(\dot{\mathbf{p}}_{i})_{eci}^{T}\mathbf{A}_{i}-(\dot{\mathbf{p}}_{1})_{eci}^{T}.
	\end{align}
\end{subequations}
	
	Note that (\ref{dii}) and (\ref{ddii}) constitute a set of linear equations with unknowns $(\mathbf{s}_{1})_{eci} $, $d_{1}$, and $\dot{d}_{1}$. Define $ \mathbf{z}_{1}=[(\mathbf{s}_{1})_{eci}^{T},d_{1},\dot{d}_{1}]^{T} $. The error vector $ \boldsymbol{\epsilon} $ in (\ref{h1}) can be then rewritten as
	\begin{align}\label{epsilon}
	\boldsymbol{\epsilon}=\mathbf{b}_{1}-\mathbf{G}_{1}\mathbf{z}_{1},
	\end{align}
	where 
	\begin{equation}
	\mathbf{b}_{1}=\left[\begin{array}{c}{\widetilde d_{2,1}^{2}} \\ {\widetilde d_{3,1}^{2}} \\ {\vdots} \\ {\widetilde d_{M,1}^{2}} \\ {2 \widetilde d_{2, 1} \widetilde{\dot{d}}_{2, 1}} \\ {2\widetilde d_{3, 1} \widetilde{\dot{d}}_{3, 1}} \\ {\vdots} \\ {2\widetilde d_{M, 1} \widetilde{\dot{d}}_{M, 1}}\end{array}\right],
	\end{equation}
	\begin{equation}
	\begin{aligned}
	\mathbf{G}_{1}=-2\cdot\left[\begin{array}{ccc}{\mathbf{w}_{2}} & {\tilde{d}_{2,1}} & {0} \\ {\mathbf{w}_{3}} & {\tilde{d}_{3,1}} & {0} \\ {\vdots} & {\vdots} & {\vdots} \\ {\mathbf{w}_{M}} & {\tilde{d}_{M, 1}} & {0} \\ {\mathbf{v}}_{2} & {\tilde{\dot{d}}_{2, 1}} & {\tilde{d}_{2, 1}} \\ {\mathbf{v}}_{3} & {\tilde{\dot{d}}_{3, 1}} & {\tilde{d}_{3, 1}} \\ {\vdots} & {\vdots} & {\vdots} \\ {\mathbf{v}}_{M} & {\tilde{\dot{d}}_{M, 1}} & {\tilde{d}_{M,1}}\end{array}\right].
	\end{aligned}
	\end{equation}
	
	The elements in $ \mathbf{z}_{1} $ are related to each other. $ d_{1} $ and $ \dot{d}_{1} $ in $ \mathbf{z}_{1} $ are related to $ (\mathbf{s}_{1})_{eci} $ through the nonlinear relationships (\ref{di}) and (\ref{ddi}) at $ i=1 $. In addition, as the satellite moves in a definite circular orbit, (\ref{opc}) is satisfied and its equivalent expression in ECEF can be written as
	\begin{align}
		(\mathbf{s}_{1})_{eci}^{T}(\mathbf{s}_{1})_{eci}=r^{2},\quad\mathbf{g}^{T}(\mathbf{s}_{1})_{eci}=0,
	\end{align}
	where $ \mathbf{g}^{T}=[\mathbf{E}^{opc}_{eci}]_{3,:} $.
	
	Like the location estimation problem in (\ref{obj1}), with the above constraints, the ephemeris estimation problem can be written as 
	\begin{align}\label{obj2}
	\mathop{\mathrm{minimize}}\limits_{\mathbf{z}_{2}}\quad & (\mathbf{b}_{2}-\mathbf{G}_{1}\mathbf{z}_{2})^{T}\boldsymbol \Psi^{-1}(\mathbf{b}_{2}-\mathbf{G}_{1}\mathbf{z}_{2}), \nonumber\\
	\mathrm{subject}\ \mathrm{to} \quad& \mathbf{z}_{2}^{T}\mathbf{C}_{5}\mathbf{z}_{2}+2\mathbf{q}_{2}^{T}\mathbf{z}_{2}-\rho_{2}=0, \nonumber\\
	& \mathbf{q}_{3}^{T}\mathbf{z}_{2}-\rho_{3}=0, \nonumber\\
	&\mathbf{z}_{2}^{T}\mathbf{C}_{6}\mathbf{z}_{2}=0,\nonumber\\
	&\mathbf{z}_{2}^{T}\mathbf{C}_{7}\mathbf{z}_{2}+\mathbf{q}_{4}^{T}\mathbf{z}_{2}=0,
	\end{align} 
	where 
	\begin{subequations}
		\begin{align}
		\mathbf{z}_{2}&=\mathbf{z}_{1}-\widetilde{\mathbf{r}}_{2},\\
		\widetilde{\mathbf{r}}_{2}&=((\mathbf{p}_{1})_{eci}^{T},0,0)^{T},\\
		\widetilde{\mathbf{r}}_{3}&=((\dot{\mathbf{p}}_{1})_{eci}^{T},0,0)^{T},\\
		\mathbf{b}_{2}&=\mathbf{b}_{1}-\mathbf{G}_{1}\widetilde{\mathbf{r}}_{2},\\
		\mathbf{C}_{5}&=\mathrm{diag}{[1,1,1,0,0]},\\
		\mathbf{q}_{2}&=\mathbf{C}_{5}\widetilde{\mathbf{r}}_{2},\\
		\mathbf{q}_{3}&=[\mathbf{g}^{T},0,0]^{T},\\
		\mathbf{q}_{4}&=\mathbf{C}_{7}\widetilde{\mathbf{r}}_{2}+\widetilde{\mathbf{r}}_{3},\\ 
		\rho_{2}&=r^{2}-(\mathbf{p}_{1})^{T}_{eci}(\mathbf{p}_{1})_{eci},\\
		\rho_{3}&=-\mathbf{g}^{T}(\mathbf{p}_{1})_{eci},\\
		\mathbf{C}_{6}&=\mathrm{diag}{[1,1,1,-1,0]},\\
		\mathbf{C}_{7}&=\left(\begin{array}{ccc}{\bPhi} & {\mathbf{0}_{3 \times 1}} & {\mathbf{0}_{3 \times 1}} \\ {\mathbf{0}_{1 \times 3}} & {0} & {-1} \\ {\mathbf{0}_{1 \times 3}} & {0} & {0}\end{array}\right).
		\end{align}
	\end{subequations}
	\subsection{The Ephemeris Estimation Algorithm}
	The ephemeris estimation problem in (\ref{obj2}) is a quadratic programming with equality constraints. As the problems in (\ref{obj2}) and (\ref{obj1}) are essentially the same kind of optimization problem, we employ the CWLS method to solve the problem in (\ref{obj2}). The procedure of the ephemeris estimation algorithm with the downlink synchronization signals is given as follows.
	
	\begin{itemize}
		\item[1)] Initialize $ k=0 $, $ \mathbf{B}^{0}=\mathbf{I} $, $ \dot{\mathbf{B}}^{0}=\mathbf{0} $, $ \boldsymbol{\Psi}^{0} $ by (\ref{psi}), and $ \mathbf{\widehat z}_{2}^{0}=(\mathbf{G}_{1}^{T}(\boldsymbol\Psi^{0})^{-1}\mathbf{G}_{1})^{-1}\mathbf{G}_{1}^{T}(\boldsymbol\Psi^{0})^{-1}\mathbf{b}_{2} $.
		\item[2)] Set $ k=k+1 $, and the approximate quadratic programming with linear equallity constraints based on (\ref{obj2}) can be written as
		\begin{align}\label{obj3}
		\mathop{\mathrm{minimize}}\limits_{\mathbf{z}_{2}}\quad & (\mathbf{b}_{2}-\mathbf{G}_{1}\mathbf{z}_{2})^{T}(\boldsymbol \Psi^{k-1})^{-1}(\mathbf{b}_{2}-\mathbf{G}_{1}\mathbf{z}_{2}), \ntb
		\mathrm{subject}\ \mathrm{to} \quad& ((\mathbf{\widehat z}_{2}^{k-1})^{T}\mathbf{C}_{5}+2\mathbf{q}_{2}^{T})\mathbf{z}_{2}=\rho_{2},\ntb
		& \bq_{3}^{T}\mathbf{z}_{2}=\rho_{3},\ntb
		& (\mathbf{\widehat z}_{2}^{k-1})^{T}\mathbf{C}_{6}\mathbf{z}_{2}=0,\ntb
		& ((\mathbf{\widehat z}_{2}^{k-1})^{T}\mathbf{C}_{7}+\bq_{4}^{T})\mathbf{z}_{2}=0.
		\end{align}
		\item[3)] Solve the problem (\ref{obj3}) by:
		\begin{align}
			\mathbf{\widehat z}_{2}^{k}=(\mathbf{P}_{3}^{k-1}\mathbf{G}_{1}^{T}(\boldsymbol\Psi^{k-1})^{-1}\mathbf{G}_{1}\mathbf{P}_{3}^{k-1})^{\dagger}(\mathbf{G}_{1}^{T}(\boldsymbol\Psi^{k-1})^{-1}\mathbf{b}_{2}\nonumber\\
			-\mathbf{G}_{1}^{T}(\boldsymbol\Psi^{k-1})^{-1}\mathbf{G}_{1}\mathbf{P}_{4}^{k-1})+\mathbf{P}_{4}^{k-1},
		\end{align}
		where
		\begin{subequations}
			\begin{align}
				\mathbf{P}_{3}^{k-1}=\mathbf{I}-(\bY^{k-1})^{T}(\bY^{k-1}(\bY^{k-1})^{T})^{-1}\bY^{k-1},
			\end{align}
			\begin{align}
				\mathbf{P}_{4}^{k-1}=(\bY^{k-1})^{T}(\bY^{k-1}(\bY^{k-1})^{T})^{-1}\boldsymbol\beta_{2},
			\end{align}
			\begin{align}
				\bY^{k-1}=&[((\mathbf{\widehat z}_{2}^{k-1})^{T}\mathbf{C}_{5}+2\mathbf{q}_{2}^{T});\mathbf{q}_{3}^{T};\ntb&(\mathbf{\widehat z}_{2}^{k-1})^{T}\mathbf{C}_{6};((\mathbf{\widehat z}_{2}^{k-1})\mathbf{C}_{7}+\bq_{4}^{T})],
			\end{align}
			\begin{align}
			\boldsymbol\beta_{2}=[\rho_{2};\rho_{3};0;0].
			\end{align}	
		\end{subequations}
		\item[4)] Estimate the location of the satellite in the $ k $-th iteration:
		\begin{align}
			(\mathbf{\widehat s}_{1})_{eci}^{k}=\mathbf{\widehat z}_{2}^{k}(1:3)+(\mathbf{p}_{1})_{eci}.
		\end{align}
	    \item[5)] Update $ \mathbf{B}^{k} $, $ \dot{\mathbf{B}}^{k} $, and $ \boldsymbol{\Psi}^{k}  $ by (\ref{psi}), (\ref{Bdiag}), and (\ref{dBdiag}), respectively and go to step 2).
	\end{itemize}
	The derivation of the constrained CRLB for the satellite location $ \mathbf{s}_{1} $ follows the same procedure as that in the Section \ref{location_esti}. 
	
	\section{Timing Advance Estimation for Multi-Satellite Systems}\label{multi-satellite}  
	
	The solutions in the above sections apply to TA estimation in single-satellite networks. However, with the rapid growth on the quantity and type of satellites, the ground UEs may receive downlink synchronization signals from multiple satellites at the same time \cite{zhao19multi}. Here, we extend the problem of TA estimation to the multi-satellite case in \textit{Scenario 1}.
	
	Consider that the UE receives downlink synchronization signals from $ G $ satellites that are located at $ \mathbf{s}_{i,g}=[x_{i,g},y_{i,g},z_{i,g}]^{T} $, and move with velocities $ \dot{\mathbf{s}}_{i,g}=[\dot{x}_{i,g},\dot{y}_{i,g},\dot{z}_{i,g}]^{T} $ when launching the $ i $-th SSB, $ i=1,2,...,M $, $ g=1,2,...,G $. Taking the signal transmitted from satellite 1 corresponding to the first downlink SSB as the reference signal, TDOA $ \widetilde t_{m,1} $ and FDOA measurements $ \widetilde f_{m,1} $, $ m=2,3,..,GM $ can be represented by (\ref{equ1_1}) and (\ref{equ1_2}), respectively. Then, we adopt the location estimation algorithms in Section \ref{location_esti} to estimate the propagation delay between UE and satellites, thus TA estimation can be calculated at the UE side.
 	
\begin{figure}
 		\centering
 		\includegraphics[width=\figdoucolwid]{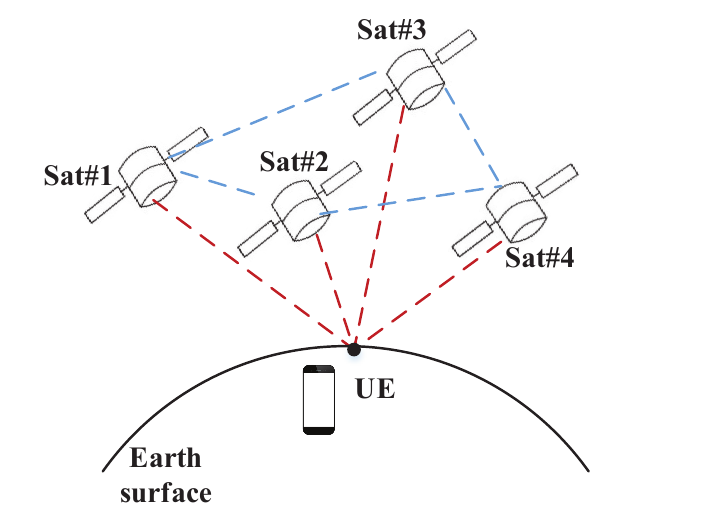}
 		\caption{Geometry of the UE in relationship to the satellites with good GDOP.}
 		\label{geometry1}
 	\end{figure}

\begin{figure}
    	\centering
    	\includegraphics[width=\figdoucolwid]{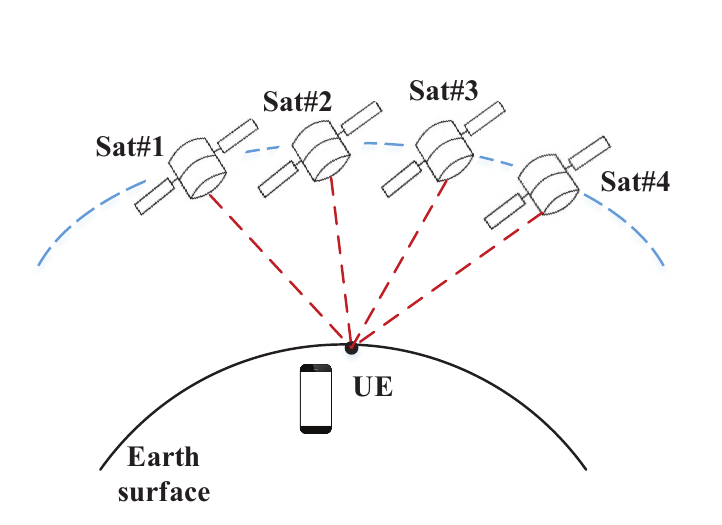}
    	\caption{Geometry of the UE in relationship to the satellites with poor GDOP.}
    	\label{geometry2}
    \end{figure}

 	The multi-satellite systems have several characteristics, e.g., more extensive coverage area, more complex connection relationships, and dynamic geometry change. \figref{geometry1} and \figref{geometry2} illustrate two different arrangements of satellites, respectively. In \figref{geometry2}, satellites are placed almost in a curvature, which limits the length of the visibility window of satellite links and thus results in poor Geometric Dilution of Precision (GDOP), while the satellites in \figref{geometry1} are distributed at well-spaced positions with a better geometry \cite{perf20zohair,GDOP09sh}. We will compare and analyze the performances of our proposed methods for TA estimation under typical situations in the following simulations.
	
	\section{Numerical Results}\label{numerical_resu}
	
	The numerical results are provided to evaluate the performance of our proposed CWLS method for estimating the TA in 5G integrated LEO SatCom. In this section, the TDOA and FDOA measurements are generated by applying the time-and-frequency synchronization algorithm in \cite{Near}. Due to the requirements of downlink synchronization defined for NTN UE in \cite{3gpp.38.811}, the signal-to-noise ratio (SNR) of synchronization in downlink is set to be -6 dB in this simulation. A multipath fading channel with system bandwidth 20 MHz and sampling frequency 30.72 MHz is adopted. In addition, the CP duration is assumed to be 0.67 ms, which corresponds to the standardized value for a subcarrier spacing of 15 kHz, and delay spread is set to be 250 ns, which is stated to cover 90 \% of the cases \cite{3gpp.38.811}.
	
	The positioning accuracy is evaluated in terms of root MSE (RMSE). For\textit{ Scenario 1} and \textit{Scenario 2}, they are defined as
	\begin{align}
	\mathrm{RMSE}_{1}=\sqrt{\frac{\sum_{l=1}^{L}\rVert\hat{\mathbf{p}}^{l}-\mathbf{p}\rVert^{2}}{L}},
	\end{align}
	\begin{align}
	\mathrm{RMSE}_{2}=\sqrt{\frac{\sum_{l=1}^{L}\rVert\hat{\mathbf{s}}^{l}_{1}-\mathbf{s}_{1}\rVert^{2}}{L}},
	\end{align}
	respectively, where $ \hat{\mathbf{p}}^{l} $ denotes the estimate of  UE position of the $ l $-th run, $ \hat{\mathbf{s}}^{l}_{1} $ denotes the estimate of satellite position of the $ l $-th run at the transmit  instant of the first SSB, and $ L=2000 $ is the number of independent runs. 
	
	TA estimation error is defined as the differential value between the actual satellite-UE distance and its estimate. Hence, the TA estimation error for the single-satellite scenario can be given by
	\begin{align}
	\Delta d=\rvert\hat d_{1}-d_{1} \rvert,
	\end{align}
	where $ \hat d_{1} $ denotes the estimate of the distance between the satellite and UE corresponding to the first downlink SSB. For the multi-satellite scenario, we consider taking satellite 1 as the reference, then the estimate of distance is calculated between satellite 1 and UE.
	
	\subsection{Single-Satellite Systems}
	
\begin{figure}
	\centering
	\includegraphics[width=\figdoucolwid]{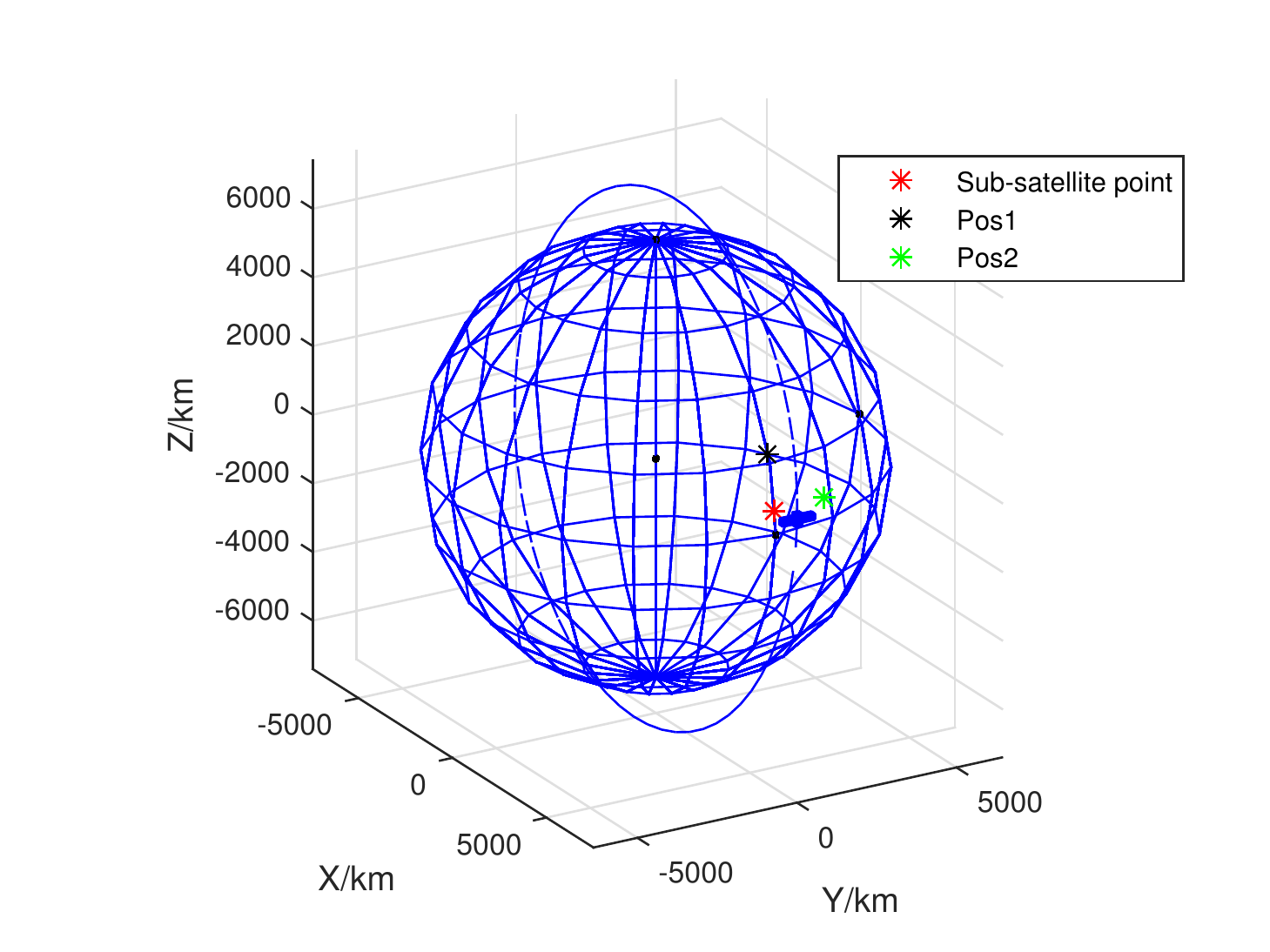}
	\caption{The distribution of three representative UE locations.}
	\label{pos3_6_8}
    \end{figure}

	\begin{table}
		\footnotesize
		\caption{Simulation Setup Parameters}\label{tb:sim_cor_par}
		\centering
		\ra{1.3}
		\begin{tabular}{LcR}
			\toprule
			Parameter & & Value\\
			\midrule
			\rowcolor{lightblue}
			Orbital altitude && 1070 km \\
			Eccentricity of the orbit && 0\\
			\rowcolor{lightblue}
			Inclination of the orbit && $85^{\circ}$ \\
			Argument of perigee of the orbit && $ 0^{\circ} $ \\
			\rowcolor{lightblue}
			Right ascension of ascending node of the orbit && $ 0^{\circ} $  \\
			Carrier frequency && 2.6 GHz  \\
			\rowcolor{lightblue}
			Half viewing angle of the satellite && $57^{\circ}$ \\
			Minimum elevation angle of the UE && $20^{\circ}$ \\
			\bottomrule
		\end{tabular}
	\end{table}
	In this simulation part, we analyze the performance of our proposed methods in single-satellite systems.
	The major simulation setup parameters for single-satellite systems are listed in Table \ref{tb:sim_cor_par}. We first select three representative UE locations in the satellite beam coverage area, which are sub-satellite point and two positions (\textit{Pos1} and \textit{Pos2}) at the region edge with lower elevation angles. \textit{Pos1} is in the direction of the sub-satellite trajectory and \textit{Pos2} is in the vertical direction of this trajectory. The detailed locations of these ground terminals are given in Table \ref{tb:sim_cor_par_1}. \figref{pos3_6_8} illustrates the distribution of these UEs.
	
	\begin{table}
		\footnotesize
		\caption{Location of the Ground Terminals}\label{tb:sim_cor_par_1}
		\centering
		\ra{1.3}
		\begin{tabular}{LcRcR}
			\toprule
			Ground terminal & & Location & &  Elevation angle\\
			\midrule
			\rowcolor{lightblue}
			Sub-satellite point && ($ 6^{\circ}$N\ ,\ $ 0^{\circ}$E) && $90^{\circ}$\\
			Pos1 && ($ 20^{\circ}$N\ ,\ $ 0^{\circ}$E) && $22^{\circ}$\\
			\rowcolor{lightblue}
			Pos2 && ($ 6^{\circ}$N\ ,\ $ 15^{\circ}$E) && $22^{\circ}$\\
			\bottomrule
		\end{tabular}
	\end{table}

\begin{figure}
		\centering
		\includegraphics[width=\figdoucolwid]{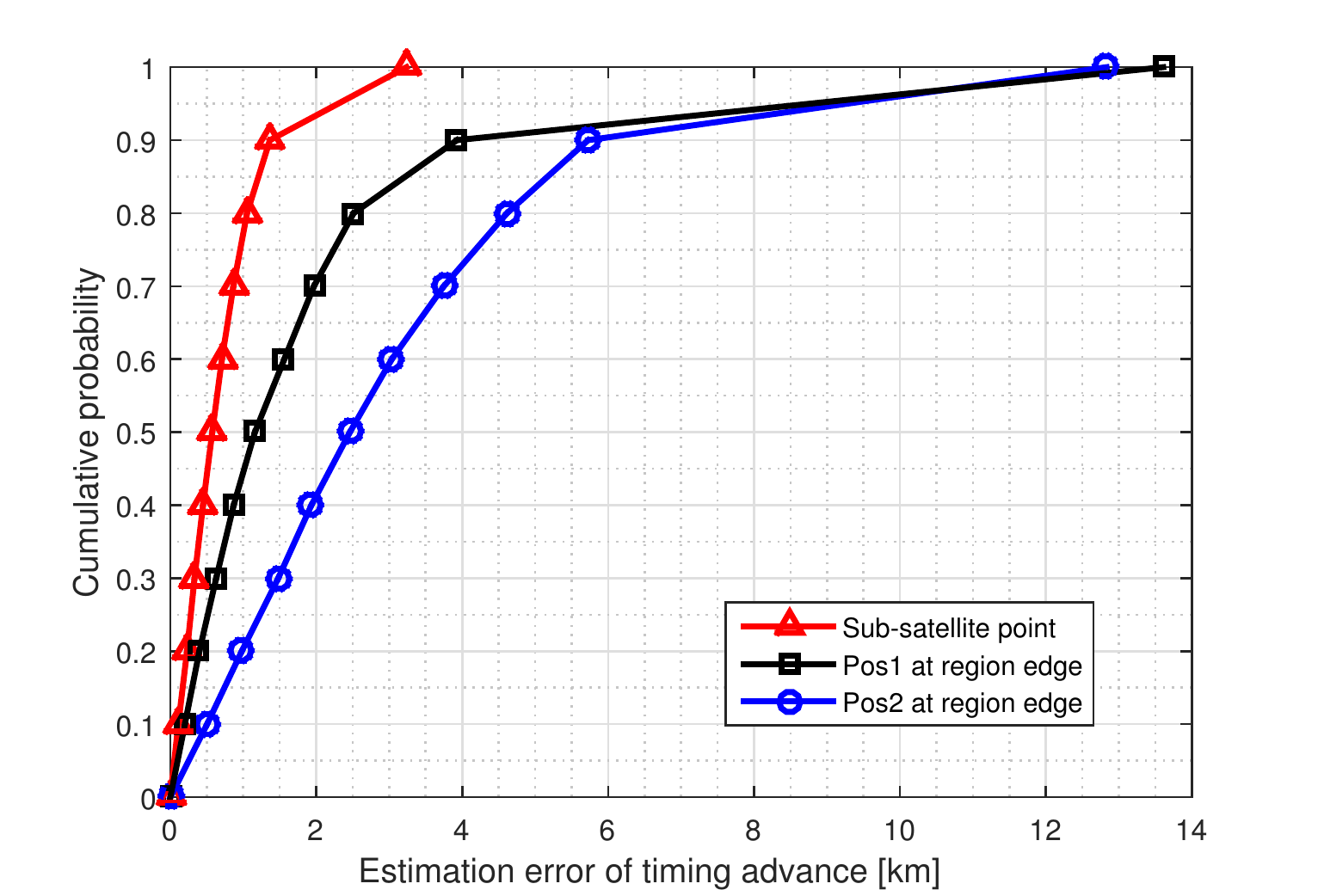}
		\caption{The cumulative probability distribution of TA estimation error in \textit{Scenario 1} (Timing-window: 12 s).}
		\label{fig1_1117}
	\end{figure}

\begin{figure}
		\centering
		\includegraphics[width=\figdoucolwid]{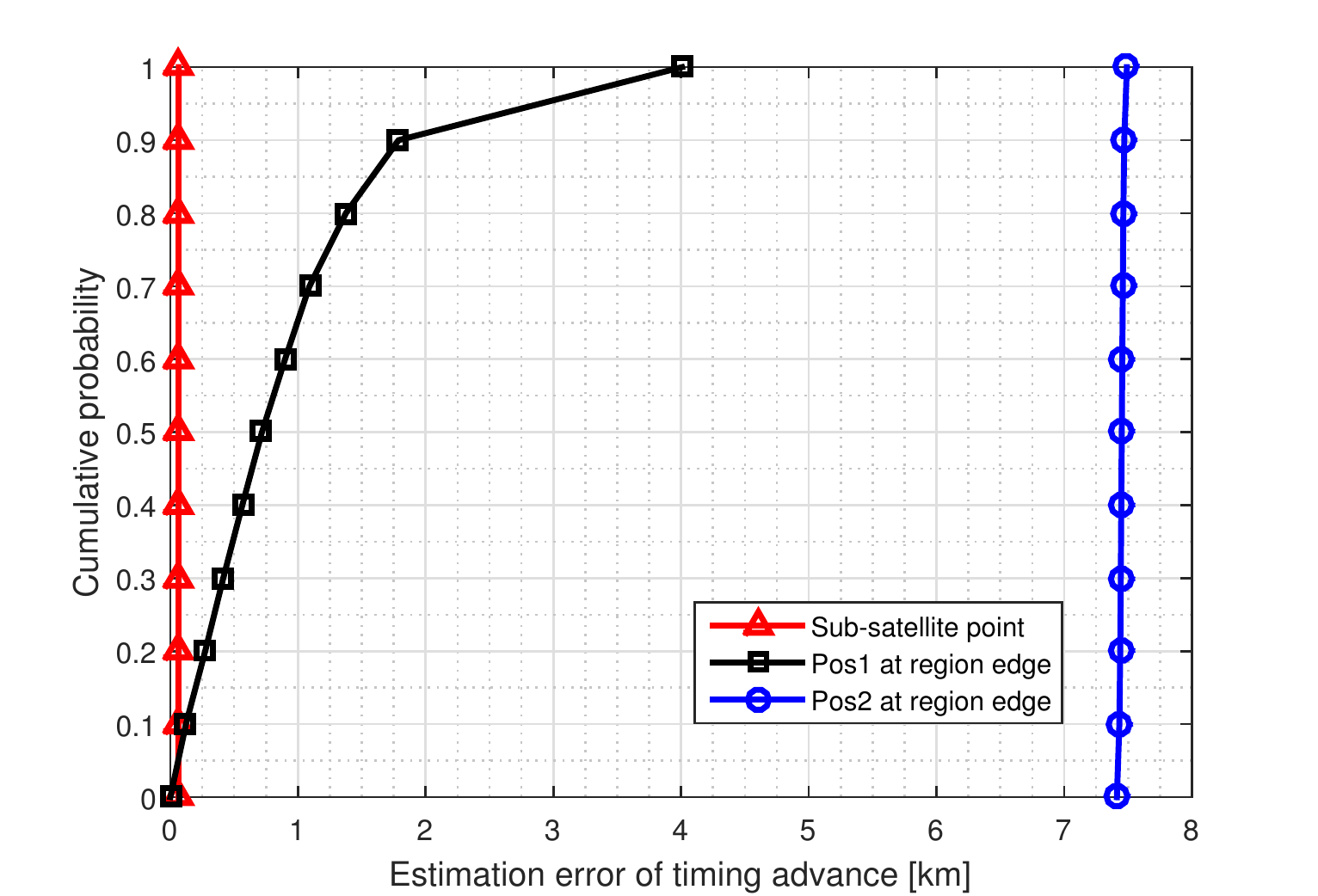}
		\caption{The cumulative probability distribution of TA estimation error in \textit{Scenario 2} (Timing-window: 2 s).}
		\label{2s3p}
	\end{figure}

	We first study the shortest possible timing-window with the guaranteed performance of TA estimation.
	Keeping the time interval between two adjacent SSBs = 20 ms, Fig. \ref{fig1_1117} and \figref{2s3p} show the cumulative probability distribution of TA estimation error of different UEs in \textit{Scenario 1} and \textit{Scenario 2}, respectively. We can observe that for a given SSB interval of 20 ms, the timing-window of 12 s can guarantee that the TA estimation offsets of all uplink frames are no more than 14 km in \textit{Scenario 1}, and the length of timing-window can be reduced to 2 s in \textit{Scenario 2}. Considering the CP type and the delay spread value, the timing-window can ensure that the TA estimation offsets of all uplink frames fall with the range of one CP. As the satellite location to be estimated in \textit{Scenario 2} is constrained by a definite orbit while the possible estimates of UE location in \textit{Scenario 1} follow the spherical equation of the earth, the performance of the TA estimation algorithm is better in \textit{Scenario 2}.
	 
	\begin{table}
		\footnotesize
		\caption{Average Runtime (in Sec.) for Sub-satellite Point in Scenario 1 (Timing-window: 10 s; SSB Interval: 20 ms) }\label{tb:sim_cor_par_2}
		\centering
		\ra{1.3}
		\begin{tabular}{LcR}
			\toprule
			Method & & Average time\\
			\midrule
			\rowcolor{lightblue}
			Quadratic penalty method &&  53.10  \\
			Iterative CWLS method &&  6.24  \\
			\bottomrule
		\end{tabular}
	\end{table}
	
	\tabref{tb:sim_cor_par_2} compares the runtimes of quadratic penalty and iterative CWLS methods. In this simulation, we set timing-window = 10 s and SSB interval = 20 ms. The simulation was conducted by MATLAB on a desktop computer with Inter i7-8700 processor and memory capacity 16 GB. We can observe that the average computation time of the quadratic penalty method is significantly larger than that of the iterative CWLS method.

\begin{figure}
		\centering
		\includegraphics[width=\figdoucolwid]{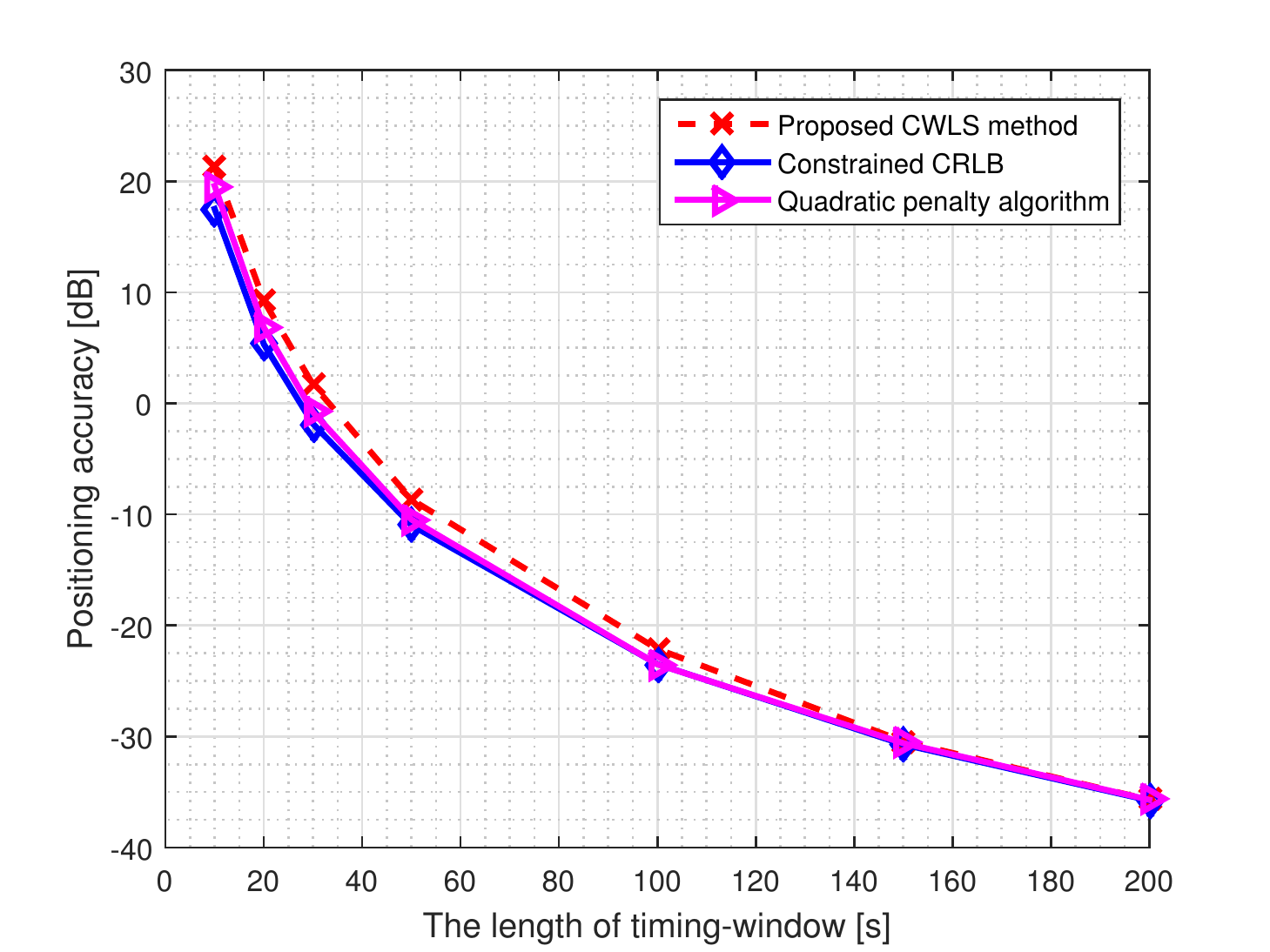}
		\caption{Comparison of positioning accuracy with CRLB in \textit{Scenario 1}  (Terminal: \textit{Pos2}; SSB interval: 20 ms).}
		\label{fig2_1117}
	\end{figure}

\begin{figure}
		\centering
		\includegraphics[width=\figdoucolwid]{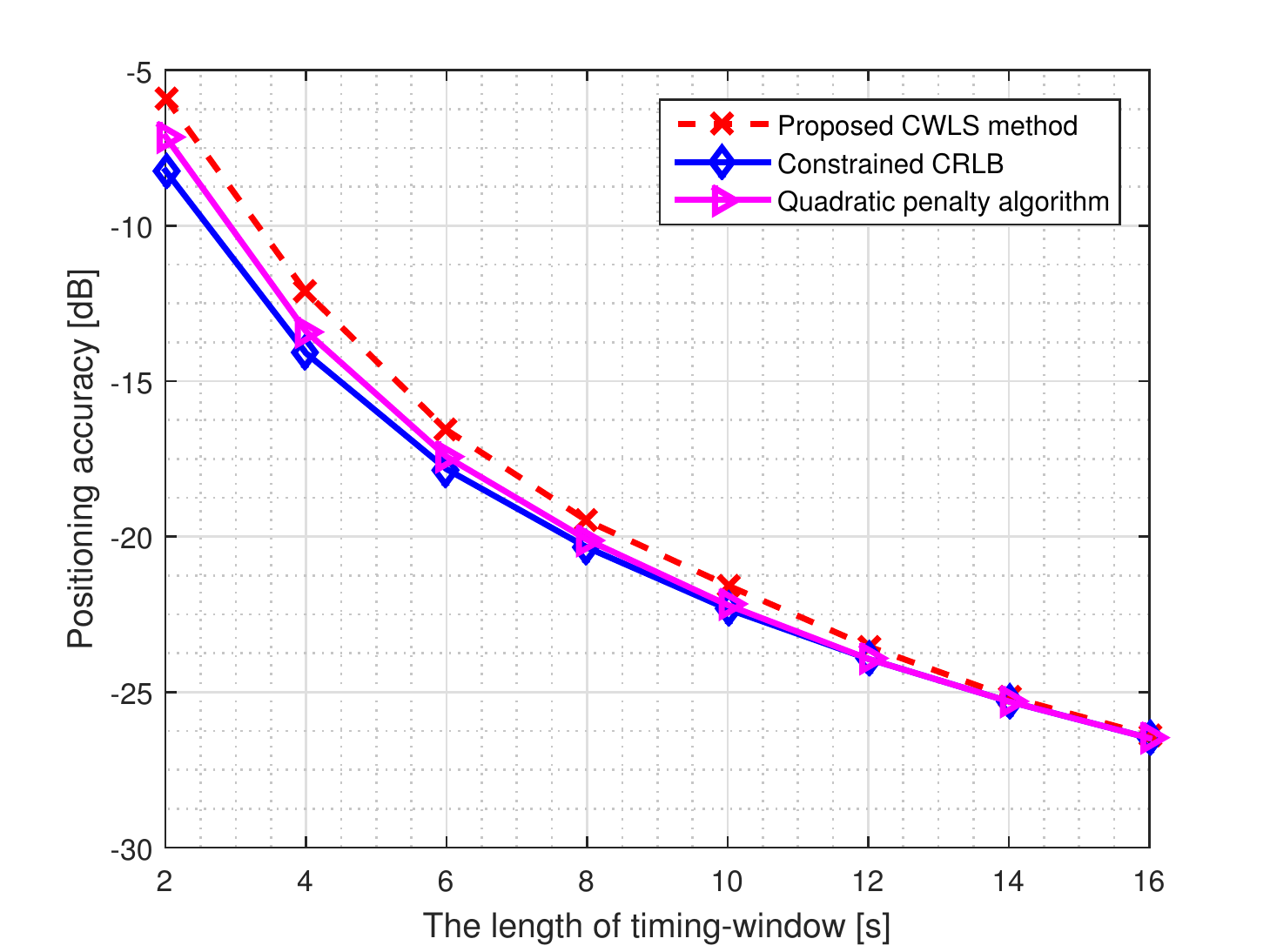}
		\caption{Comparison of positioning accuracy with CRLB in \textit{Scenario 2}  (Terminal: Sub-satellite point; SSB interval: 20 ms).}
		\label{fig2_0522}
	\end{figure}

	Fig. \ref{fig2_1117} and \figref{fig2_0522} compare the MSE of our proposed CWLS method and the quadratic penalty algorithm with the constrained CRLB in \textit{Scenario 1} and \textit{Scenario 2}, respectively. We observe that the proposed CWLS method approaches the CRLB effectively, and that the performance of the proposed CWLS method has little degradation compared with the quadratic penalty algorithm.
	
\begin{figure}
		\centering
		\includegraphics[width=\figdoucolwid]{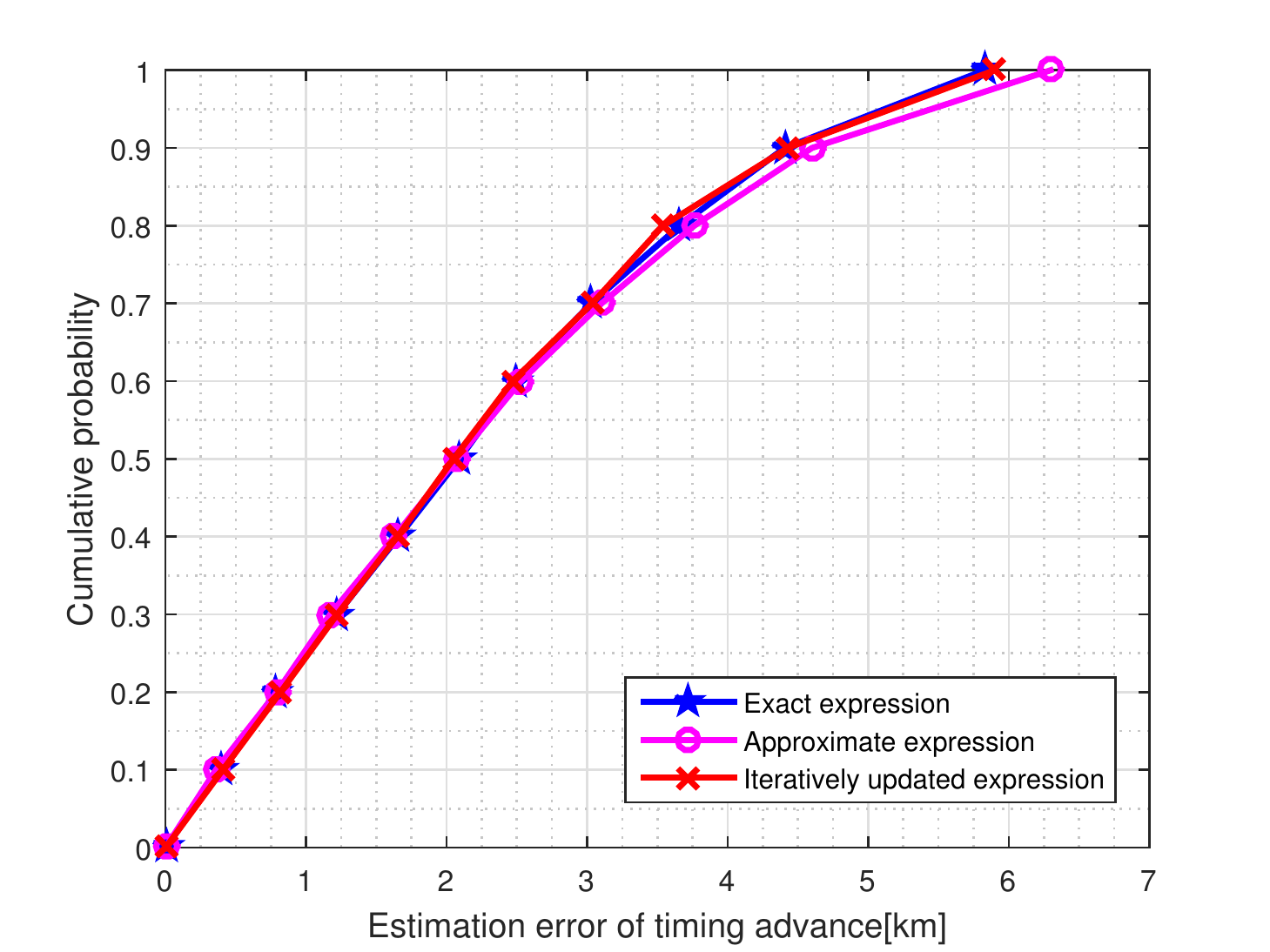}
		\caption{The cumulative probability distribution of TA estimation error in \textit{Scenario 1} with the weighting matrix taking the exact, approximate without the update, and iteratively updated values (Timing-window: 12 s; SSB interval: 20 ms). }
		\label{fig1}
	\end{figure}

\begin{figure}
		\centering
		\includegraphics[width=\figdoucolwid]{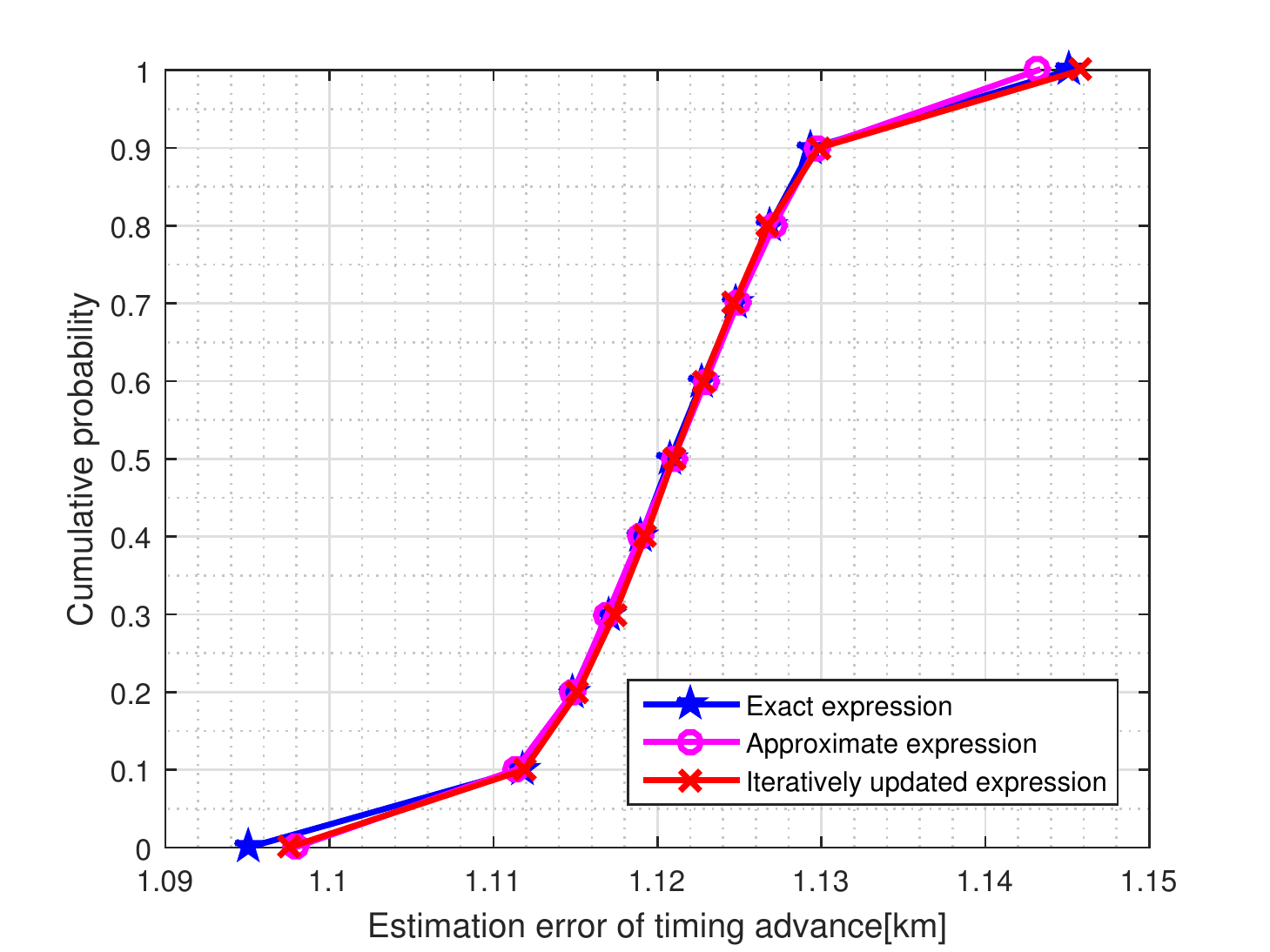}
		\caption{The cumulative probability distribution of TA estimation error \textit{in Scenario 2} with the weighting matrix taking the exact, approximate without the update, and iteratively updated values (Timing-window: 2 s; SSB interval: 20 ms).}
		\label{fig2}
	\end{figure}

	In addition, in Sections \uppercase\expandafter{\romannumeral2} and \uppercase\expandafter{\romannumeral3}, the approximation $ \mathbf{B}\approx2d^{0}\mathbf{I} $, $ \dot{\mathbf{B}}\approx\mathbf{0} $, and their iteratively updated values are applied to calculate the weighting matrix $ \boldsymbol{\Psi} $. To verify the accuracy of the approximations, we compare the performances of our proposed method with the weighting matrix taking the exact, approximate without the update, and iteratively updated values. Fig. \ref{fig1} and Fig. \ref{fig2} show the comparisons of the cumulative probability distribution of TA estimation error in \textit{Scenario 1} and \textit{Scenario 2}, respectively. We can observe that in both scenarios, our proposed methods of TA estimation with weighting matrix taking the exact and iteratively updated values have almost the same performance. As the timing window in \textit{Scenario 1} is larger than that in \textit{Scenario 2}, the iteratively updated expression provides slightly better TA estimation results than the approximated values in \textit{Scenario 1}, but no significant improvement is observed in \textit{Scenario 2}. Hence, for the short-time random access procedure, the approximated expressions without the update are sufficient, and the iteratively updated expressions are preferred for longer initial access time.
	
	\subsection{Multi-Satellite Systems}
	
\begin{figure}
 		\centering
 		\includegraphics[width=\figdoucolwid]{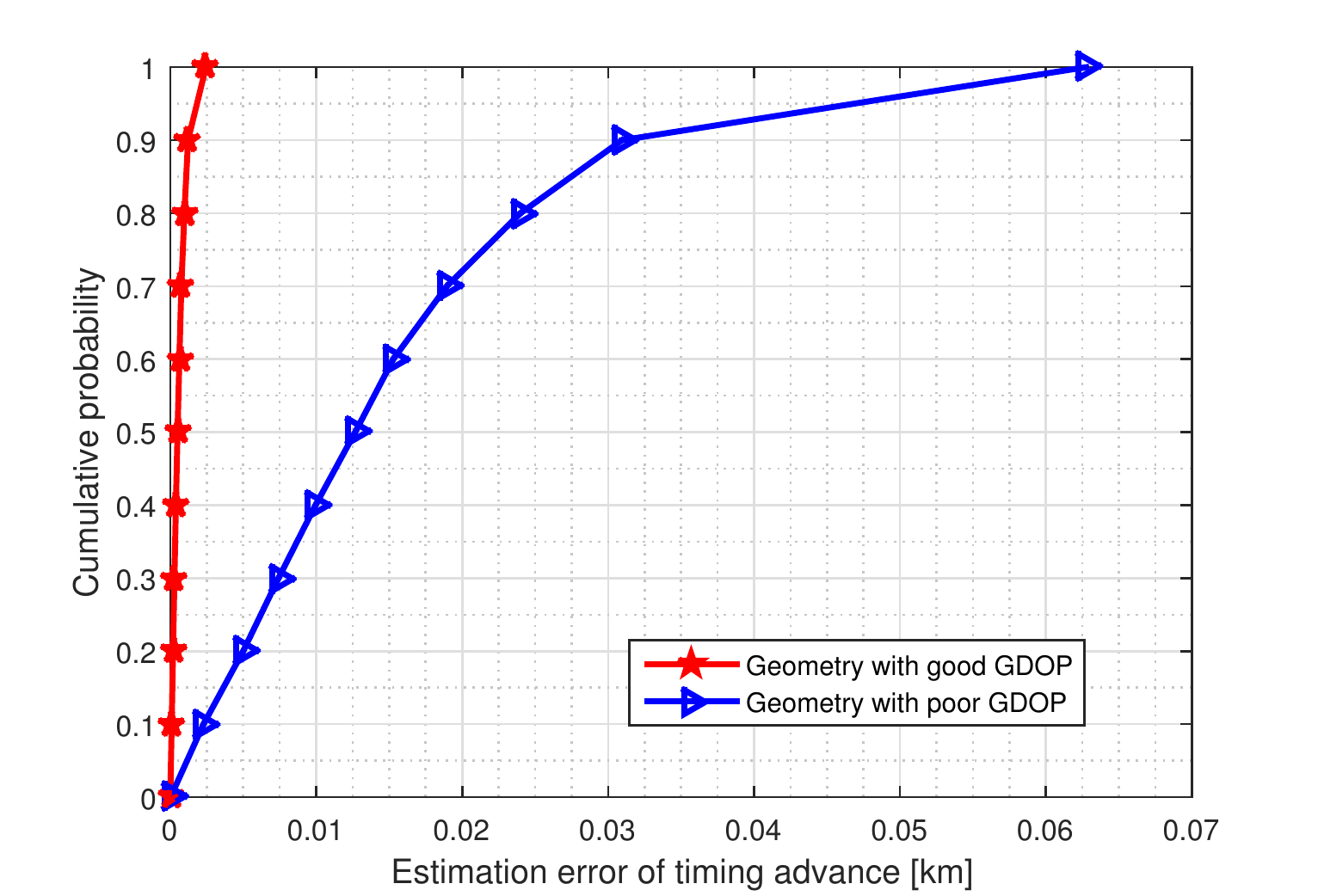}
 		\caption{The cumulative probability distribution of TA estimation error for multi-satellite systems in \textit{Scenario 1}  (Timing-window: 1 s).}
 		\label{fig3_0522}
 	\end{figure}

	The performance of our proposed CWLS methods for multi-satellite systems in \textit{Scenario 1} is discussed in this section. The number of satellites is set to be four and two scenarios with different geometries of satellites are presented. In the first scenario with good GDOP, Sat1 and Sat3 are in the same orbit, while Sat2 and Sat4 are in the same orbit. The right ascensions of ascending node of these two orbits are $ 0^{\circ} $ and $ 20^{\circ} $, respectively. The satellites in the second scenario with bad GDOP are all in the same orbit and its right ascension of ascending node is $ 0^{\circ} $. The other orbital parameters are the same as those in Table \ref{tb:sim_cor_par}. \figref{fig3_0522} shows the cumulative probability distribution of TA estimation under these two different arrangements of satellites. We can observe that the timing-window of 1s can guarantee that the TA estimation offsets of all uplink frames to fall within the range of one CP, and the performance of our proposed method for TA estimation under geometry with good GDOP is better.
 	
\begin{figure}
    	\centering
    	\includegraphics[width=\figdoucolwid]{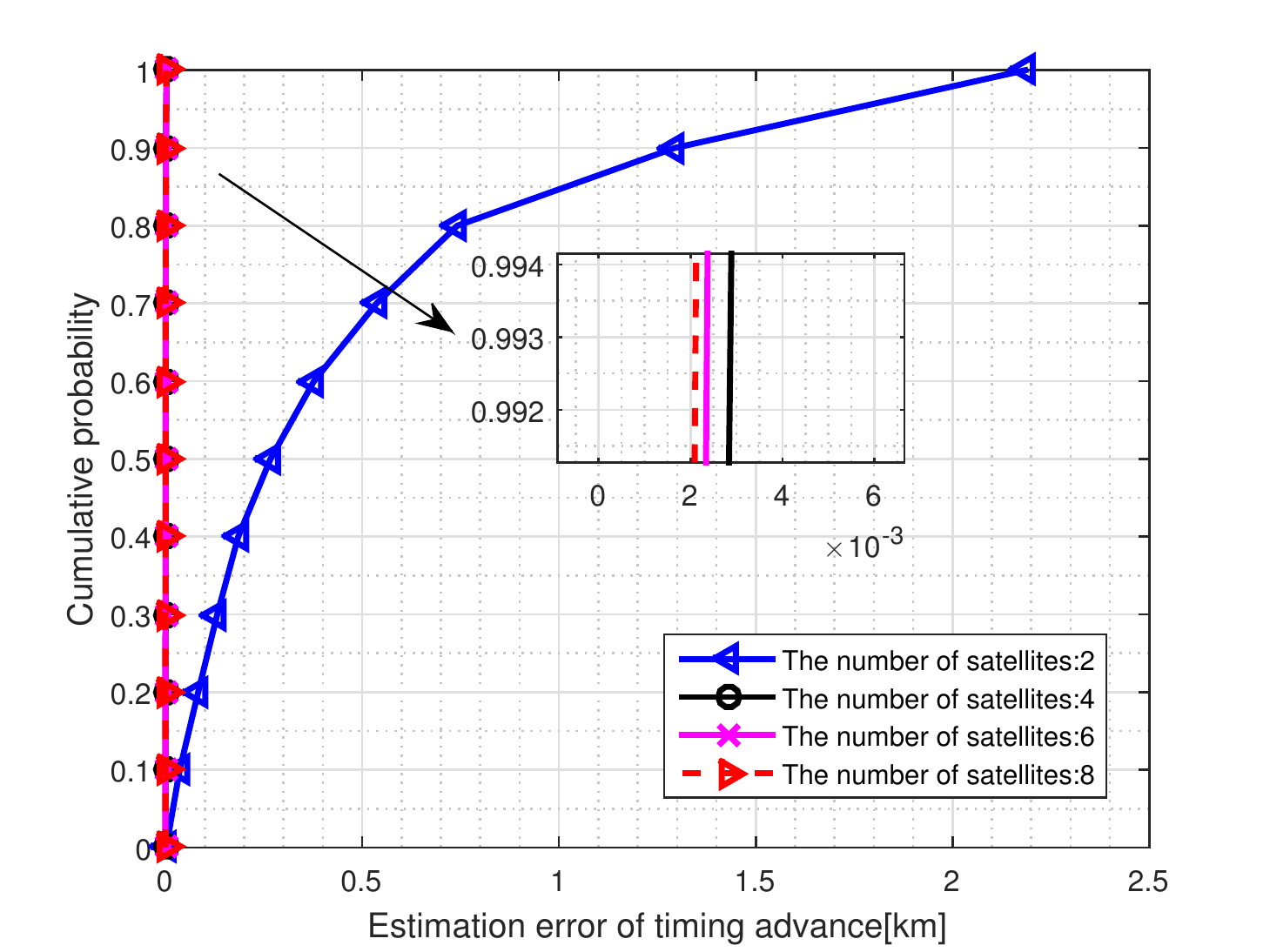}
    	\caption{The cumulative probability of TA estimation error in \textit{Scenario 1} with different number of LEO satellites (Terminal: Pos 2; SSB interval: 20 ms; timing-window: 1 s).}
    	\label{fig1_0204}
    \end{figure}

    \figref{fig1_0204} compares the cumulative probability of TA estimation error with different number of LEOs. In our simulation, the numbers of LEO satellites vary from 2, 4, 6, to 8. Except for right ascension of ascending node, the other orbital parameters are set to be the same as those in Table \ref{tb:sim_cor_par}. We can observe that the timing-window of 1 s can guarantee that the TA estimation offsets to fall within the range of one CP, and that the performances of our proposed method for TA estimation with 4, 6, and 8 satellites are significantly better than that with 2 satellites.
    
	\section{Conclusion}\label{conclusion}
	
	In this paper, we proposed a new approach aided by the UE geolocation to perform uplink TA for random access in 5G integrated LEO SatCom with the TDOA and FDOA measurements acquired in the downlink timing and frequency synchronization phase, thus solving the inadaptability of the TA scheme initially designed for 5G NR system in 5G integrated LEO SatCom. We established the system model of TA estimation in two typical scenarios separately and converted the problem into UE geolocation or satellite ephemeris estimation.
	We introduced an equality-constrained quadratic optimization problem to obtain the UE geographical position or satellite ephemeris and then adopted a quadratic penalty algorithm to find the globally optimal solution of the problem. To reduce the computational complexity, we further proposed an iterative CWLS method alternatively. Numerical results showed that the proposed method can effectively reach the constrained CRLB of TA estimation and thus achieve uplink frame alignment across UEs.
	
	\begin{appendices}
		\section{Proof of Proposition 1}
		\label{appendixa}
		Since the minimum point of the problem (\ref{obj1}) must be in the feasible set $\mathcal{D}$, the necessity is established. Proof of sufficiency is given as follows.
		
		Supposing ${\mathbf{u}}_{2,k}\in \mathcal{D}$, then for $\forall \mathbf{u}_{2}\in \mathcal{D}$, we obtain
		\begin{align}\label{A1}
		\alpha(\mathbf{u}_{2,k})&=0,\ntb
		g(\mathbf{u}_{2,k})&=F(\mathbf{u}_{2,k};\mu_{k}).
		\end{align}
		Since $\mathbf{u}_{2,k}$ is the minimum point of $F$, it leads to the following inequality
		\begin{align}\label{A2}
		F(\mathbf{u}_{2,k};\mu_{k})\leq F(\mathbf{u}_{2};\mu_{k}).
		\end{align}
		Since $\alpha(\mathbf{u}_{2})=0$, we then obtain
		\begin{align}\label{A3}
		F(\mathbf{u}_{2};\mu_{k})=g(\mathbf{u}_{2}).
		\end{align}
		From (\ref{A1}), (\ref{A2}) and (\ref{A3}), we can observe that
		\begin{align}
		g(\mathbf{u}_{2,k})\leq g(\mathbf{u}_{2}).
		\end{align}
		This concludes the proof.

		\section{Proof of Proposition 2}
		\label{appendixb}
		Supposing $\bar{\mathbf{u}}_{2}$ is a global solution of (\ref{obj1}), we obtain
		\begin{equation}
		\begin{aligned}
		g(\bar{\mathbf{u}}_{2})\leq g(\mathbf{u}_{2})\quad \mathrm{for}\ \mathrm{all}\ \mathbf{u}_{2}\ \mathrm{with}\ c_{i}(\mathbf{u}_{2})=0,\\ i=1,2,3. 
		\end{aligned}
		\end{equation}
		Since $\mathbf{u}_{2,k}$ minimizes $F(\mathbf{u}_{2};\mu_{k})$, we have $F(\mathbf{u}_{2,k};\mu_{k})\leq F(\bar{\mathbf{u}}_{2};\mu_{k})$, which leads to the following equality
		\begin{equation}
		\begin{aligned}\label{u2}
		g(\mathbf{u}_{2,k})+& \mu_{k}\sum_{i=1}^{3} c_{i}^{2}(\mathbf{u}_{2,k})\leq
		\\&g(\bar{\mathbf{u}}_{2})+\mu_{k}\sum_{i=1}^{3}c_{i}^{2}(\bar{\mathbf{u}}_{2})=g(\bar{\mathbf{u}}_{2}).
		\end{aligned}
		\end{equation} 
		By rearranging (\ref{u2}), we obtain
		\begin{align}\label{A7}
		\sum_{i=1}^{3} c_{i}^{2}(\mathbf{u}_{2,k})\leq \frac{1}{\mu_{k}}[g(\bar{\mathbf{u}}_{2})-g(\mathbf{u}_{2,k})].
		\end{align}
		Suppose that $\mathbf{u}_{2}^{*}$ is a limit point of the sequence $\{\mathbf{u}_{2,k}\}$, i.e.,
		\begin{align}
		\lim\limits_{k\to \infty}\ \mathbf{u}_{2,k}=\mathbf{u}_{2}^{*}.
		\end{align}
		By taking the limit as $k\to \infty$ on both sides of (\ref{A7}), we obtain
		\begin{equation}
		\begin{aligned}
		\sum_{i=1}^{3} c_{i}^{2}(\mathbf{u}_{2}^{*})=&\lim\limits_{k\to \infty}\ \sum_{i=1}^{3} c_{i}^{2}(\mathbf{u}_{2,k})\\ \leq &\lim\limits_{k\to \infty}\ \frac{1}{\mu_{k}}[g(\bar{\mathbf{u}}_{2})-g(\mathbf{u}_{2,k})]=0, 
		\end{aligned}
		\end{equation}
		thus we have that $c_{i}(\mathbf{u}_{2}^{*})=0$, i.e., $\mathbf{u}_{2}^{*}$ is a feasible point. By taking the limit as $k\to \infty$ and considering the nonnegativity of $\mu_{k}$, we obtain
		\begin{align}
		g(\mathbf{u}_{2}^{*})\leq g(\mathbf{u}_{2}^{*})+\lim\limits_{k\to \infty}\ \mu_{k}\sum_{i=1}^{3} c_{i}^{2}(\mathbf{u}_{2,k})\leq g(\bar{\mathbf{u}}_{2}).
		\end{align}
		Since $\mathbf{u}_{2}^{*}$ is feasible and its objective value is no larger than that of the global minimizer $\bar{\mathbf{u}}_{2}$, we can conclude that $\mathbf{u}_{2}^{*}$ is also a globally optimal solution. This concludes the proof.
		
		\section{Derivation of $ \mathbf{A}_{i} $ in (\ref{si})}
		\label{appendixc}
		From (\ref{opc}), the tranformation between $ (\mathbf{s}_{i})_{opc} $ and $ (\mathbf{s}_{1})_{opc} $ can be described as
		\begin{align}
			(\mathbf{s}_{i})_{opc}&=\mathbf{R_{z}}(-(\alpha_{i}-\alpha_{1}))(\mathbf{s}_{1})_{opc}\nonumber\\
			&=\mathbf{R_{z}}(-(i-1)\cdot n'\cdot T)(\mathbf{s}_{1})_{opc},
		\end{align}
		where $ T $ denotes the time interval between  adjacent SSBs, and $ n' $ denotes the mean motion satisfying
		\begin{align}
			n'=\sqrt{\dfrac{\mu'}{r^{3}}}.
		\end{align}
		Define $ \mathbf{E}^{eci}_{opc}=\mathbf{R_{z}}(-\Omega)\mathbf{R_{x}}(-\vartheta)\mathbf{R_{z}}(-\varphi)$. With the transformation matrix in (\ref{sit}), it further has
		\begin{align}\label{sit2}
			(\mathbf{s}_{i})_{eci}&=\mathbf{E}^{eci}_{opc}(\mathbf{s}_{i})_{opc}\nonumber\\
			&=\mathbf{E}^{eci}_{opc}\mathbf{R_{z}}(-(i-1)\cdot n'\cdot T)(\mathbf{s}_{1})_{opc}.
		\end{align}
		At $ i=1 $, (\ref{sit2}) can be rewritten as
		\begin{align}
		    (\mathbf{s}_{1})_{opc}=(\mathbf{E}^{eci}_{opc})^{-1}(\mathbf{s}_{1})_{eci}.
		\end{align}
	    Thus, the transformation matrix $ \mathbf{A}_{i} $ in (\ref{si}) is given by
	    \begin{align}
	    	\mathbf{A}_{i}=\mathbf{E}^{eci}_{opc}\mathbf{R_{z}}(-(i-1)\cdot n'\cdot T)(\mathbf{E}^{eci}_{opc})^{-1}.
	    \end{align}
	\end{appendices}



\bibliography{Refabrv,References,mmW}

\begin{thebibliography}{10}
\providecommand{\url}[1]{#1}
\csname url@samestyle\endcsname
\providecommand{\newblock}{\relax}
\providecommand{\bibinfo}[2]{#2}
\providecommand{\BIBentrySTDinterwordspacing}{\spaceskip=0pt\relax}
\providecommand{\BIBentryALTinterwordstretchfactor}{4}
\providecommand{\BIBentryALTinterwordspacing}{\spaceskip=\fontdimen2\font plus
\BIBentryALTinterwordstretchfactor\fontdimen3\font minus
  \fontdimen4\font\relax}
\providecommand{\BIBforeignlanguage}[2]{{%
\expandafter\ifx\csname l@#1\endcsname\relax
\typeout{** WARNING: IEEEtran.bst: No hyphenation pattern has been}%
\typeout{** loaded for the language `#1'. Using the pattern for}%
\typeout{** the default language instead.}%
\else
\language=\csname l@#1\endcsname
\fi
#2}}
\providecommand{\BIBdecl}{\relax}
\BIBdecl

\bibitem{Integ}
O.~Kodheli, A.~Guidotti, and A.~Vanelli-Coralli, ``{Integration of satellites
  in 5G through LEO constellations},'' in \emph{Proc. IEEE GLOBECOM},
  Singapore, 2017, pp. 1--6.

\bibitem{Satellite-enabled}
A.~Guidotti, A.~Vanelli-Coralli, M.~Caus, J.~Bas, G.~Colavolpe, T.~Foggi,
  S.~Cioni, A.~Modenini, and D.~Tarchi, ``{Satellite-enabled LTE systems in LEO
  constellations},'' in \emph{Proc. IEEE ICC}, Paris, France, 2017, pp.
  876--881.

\bibitem{Energy}
Y.~Ruan, Y.~Li, C.-X. Wang, and R.~Zhang, ``{Energy efficient adaptive
  transmissions in integrated satellite-terrestrial networks with SER
  constraints},'' \emph{{IEEE} Trans. Wireless Commun.}, vol.~17, no.~1, pp.
  210--222, Jan. 2018.

\bibitem{Satellite-5G}
G.~Giambene, S.~Kota, and P.~Pillai, ``{Satellite-5G integration: A network
  perspective},'' \emph{{IEEE} Netw.}, vol.~32, no.~5, pp. 25--31, Sep./Oct.
  2018.

\bibitem{kato19optim}
N.~Kato, Z.~M. Fadlullah, F.~Tang, B.~Mao, S.~Tani, A.~Okamura, and J.~Liu,
  ``{Optimizing space-air-ground integrated networks by artificial
  intelligence},'' \emph{{IEEE} Wireless Commun.}, vol.~26, no.~4, pp.
  140--147, Aug. 2019.

\bibitem{Jia18space}
J.~Liu, Y.~Shi, Z.~M. Fadlullah, and N.~Kato, ``{Space-air-ground integrated
  network: A survey},'' \emph{{IEEE} Commun. Surveys Tuts.}, vol.~20, no.~4,
  pp. 2714--2741, {Fourth Quarter} 2018.

\bibitem{3gpp.38.811}
{3GPP TR 38.811 V15.2.0}, ``{3rd Generation Partnership Project; Technical
  Specification Group Radio Access Network; Study on New Radio (NR) to Support
  Non Terrestrial Networks (Release 15)},'' Tech. Rep., Sep. 2019.

\bibitem{ASurvey}
A.~Gupta and R.~K. Jha, ``{A survey of 5G network: Architecture and emerging
  technologies},'' \emph{{IEEE Access}}, vol.~3, pp. 1206--1232, Jul. 2015.

\bibitem{robust18wang}
W.~Wang, A.~Liu, Q.~Zhang, L.~You, X.~Q. Gao, and G.~Zheng, ``{Robust
  multigroup multicast transmission for frame-based multi-beam satellite
  systems},'' \emph{{IEEE Access}}, vol.~6, pp. 46\,074--46\,083, Aug. 2018.

\bibitem{outa19you}
L.~You, A.~Liu, W.~Wang, and X.~Q. Gao, ``{Outage constrained robust multigroup
  multicast beamforming for multi-beam satellite communication systems},''
  \emph{{IEEE} Wireless Commun. Lett.}, vol.~8, no.~2, pp. 352--355, Apr. 2019.

\bibitem{massive20you}
L.~You, K.-X. Li, J.~Wang, X.~Q. Gao, X.-G. Xia, and B.~Ottersten, ``{Massive
  MIMO transmission for LEO satellite communications},'' \emph{{IEEE} J. Sel.
  Areas Commun.}, vol.~38, no.~8, pp. 1851--1865, Aug. 2020.

\bibitem{sat18kap}
A.~{Kapovits}, M.~{Corici}, I.~{Gheorghe-Pop}, A.~{Gavras}, F.~{Burkhardt},
  T.~{Schlichter}, and S.~{Covaci}, ``Satellite communications integration with
  terrestrial networks,'' \emph{China Commun.}, vol.~15, no.~8, pp. 22--38,
  Aug. 2018.

\bibitem{Architectures}
A.~Guidotti, A.~Vanelli-Coralli, M.~Conti, S.~Andrenacci, S.~Chatzinotas,
  N.~Maturo, B.~Evans, A.~Awoseyila, A.~Ugolini, T.~Foggi, L.~Gaudio,
  N.~Alagha, and S.~Cioni, ``{Architectures and key technical challenges for 5G
  systems and incorporating satellites},'' \emph{{IEEE} Trans. Veh. Technol.},
  vol.~68, no.~3, pp. 2624--2639, Mar. 2019.

\bibitem{ran18xiong}
Q.~{Xiong}, B.~{Yu}, C.~{Qian}, X.~{Li}, and C.~{Sun}, ``Random access preamble
  generation and procedure design for {5G-NR} system,'' in \emph{Proc. 2018
  IEEE Globecom Workshops}, Abu Dhabi, United Arab Emirates, 2018, pp. 1--5.

\bibitem{Kons18use}
K.~Liolis, A.~Geurtz, R.~Sperber, D.~Schulz, S.~Watts, G.~Poziopoulou,
  B.~Evans, N.~Wang, O.~Vidal, B.~T. Jou, M.~Fitch, S.~D. Sendra, P.~S.
  Khodashenas, and N.~Chuberre, ``Use cases and scenarios of {5G} integrated
  satellite-terrestrial networks for enhanced mobile broadband: {The SaT5G}
  approach,'' \emph{{Int. J. Satell. Commun. Netw.}}, vol.~37, no.~2, pp.
  91--112, May 2018.

\bibitem{sch185g}
G.~{Schreiber} and M.~{Tavares}, ``{5G} new radio physical random access
  preamble design,'' in \emph{{Proc. 2018 IEEE 5GWF}}, Silicon Valley, CA, USA,
  2018, pp. 215--220.

\bibitem{Erik18nr}
E.~Dahlman, S.~Parkvall, and J.~Sk\"{o}ld, \emph{{5G NR: The Next Generation
  Wireless Access Technology}}.\hskip 1em plus 0.5em minus 0.4em\relax Waltham,
  MA, USA: Academic Press, 2018.

\bibitem{3gpp.38.213}
{3GPP TS 38.213 V15.3.0}, ``{3rd Generation Partnership Project; Technical
  Specification Group Radio Access Network; NR; Physical Layer Procedures for
  Control (Release 15)},'' Tech. Rep., Sep. 2018.

\bibitem{ran19harri}
H.~Saarnisaari, A.~O. Laiyemo, and C.~H.~M. {de Lima}, ``{Random access process
  analysis of 5G new radio based satellite links},'' in \emph{{Proc. 2019 IEEE
  2nd 5GWF}}, Dresden, Germany, 2019, pp. 654--658.

\bibitem{ran18zhen}
L.~{Zhen}, H.~{Qin}, B.~{Song}, R.~{Ding}, X.~{Du}, and M.~{Guizani}, ``Random
  access preamble design and detection for mobile satellite communication
  systems,'' \emph{{IEEE} J. Sel. Areas Commun.}, vol.~36, no.~2, pp. 280--291,
  Feb. 2018.

\bibitem{Zhen20prea}
L.~{Zhen}, T.~{Sun}, G.~{Lu}, K.~{Yu}, and R.~{Ding}, ``Preamble design and
  detection for {5G} enabled satellite random access,'' \emph{{IEEE} Access},
  vol.~8, pp. 49\,873--49\,884, Mar. 2020.

\bibitem{Si13lte}
S.~Liu, F.~Qin, Z.~Gao, Y.~Zhang, and Y.~He, ``{LTE-satellite: Chinese proposal
  for satellite component of IMT-Advanced system},'' \emph{China Commun.},
  vol.~10, no.~10, pp. 47--64, Oct. 2013.

\bibitem{Caus20new}
M.~{Caus}, A.~I. {P\'{e}rez-Neira}, J.~{Bas}, and L.~{Blanco}, ``{New satellite
  random access preamble design based on pruned DFT-spread FBMC},''
  \emph{{IEEE} Trans. Commun.}, vol.~68, no.~7, pp. 4592--4604, Jul. 2020.

\bibitem{3gpp.38.104}
{3GPP TS 38.104 V16.3.0}, ``{3rd Generation Partnership Project; Technical
  Specification Group Radio Access Network; NR; Base Station (BS) Radio
  Transmission and Reception (Release 16)},'' Tech. Rep., Mar. 2020.

\bibitem{Enhanced}
G.~Cui, Y.~He, P.~Li, and W.~Wang, ``{Enhanced timing advanced estimation with
  symmetric Zadoff-Chu sequences for satellite systems},'' \emph{{IEEE} Commun.
  Lett.}, vol.~19, no.~5, pp. 747--750, May 2015.

\bibitem{Two-step}
C.~Li, H.~Ba, H.~Duan, Y.~Gao, and J.~Wu, ``{A two-step time delay difference
  estimation method for initial random access in satellite LTE system},'' in
  \emph{{Proc. 16th Int. Conf. Adv. Commun. Technol.}}, Pyeongchang, South
  Korea, 2014, pp. 10--13.

\bibitem{Timing}
Y.~He, G.~Cui, P.~Li, R.~Chang, and W.~Wang, ``{Timing advanced estimation
  algorithm of low complexity based on DFT spectrum analysis for satellite
  system},'' \emph{China Commun.}, vol.~12, no.~4, pp. 140--150, Apr. 2015.

\bibitem{Yu20timing}
Y.~Zhang, L.~Zhen, G.~Lu, and K.~Yu, ``{Timing advance estimation with
  robustness to frequency offset in satellite mobile communications},'' in
  \emph{{Proc. IEEE ICCC}}, Chongqing, China, 2020, pp. 917--922.

\bibitem{Prin13paul}
P.~D. Groves, \emph{{Principles of GNSS, inertial, and multisensor integrated
  navigation systems}}, 2nd~ed.\hskip 1em plus 0.5em minus 0.4em\relax Boston,
  MA, USA: Artech, 2013.

\bibitem{Sun11an}
M.~{Sun} and K.~C. {Ho}, ``An asymptotically efficient estimator for {TDOA} and
  {FDOA} positioning of multiple disjoint sources in the presence of sensor
  location uncertainties,'' \emph{{IEEE} Trans. Signal Process.}, vol.~59,
  no.~7, pp. 3434--3440, Jul. 2011.

\bibitem{Kc04an}
K.~C. {Ho} and {W. Xu}, ``An accurate algebraic solution for moving source
  location using {TDOA} and {FDOA} measurements,'' \emph{{IEEE} Trans. Signal
  Process.}, vol.~52, no.~9, pp. 2453--2463, Sep. 2004.

\bibitem{Geolocation}
K.~C. Ho and Y.~T. Chan, ``{Geolocation of a known altitude object from TDOA
  and FDOA measurements},'' \emph{{IEEE} Trans. Aerosp. Electron. Syst.},
  vol.~33, no.~3, pp. 770--782, Jul. 1997.

\bibitem{Iterative}
X.~M. Qu, L.~H. Xie, and W.~R. Tan, ``{Iterative constrained weighted least
  squares source localization using TDOA and FDOA measurements},'' \emph{{IEEE}
  Trans. Signal Process.}, vol.~65, no.~15, pp. 3990--4003, Aug. 2017.

\bibitem{Por19tech}
I.~del Portillo, B.~G. Cameron, and E.~F. Cramley, ``A technical comparison of
  three low earth orbit satellite constellation systems to provide global
  broadband,'' \emph{{Acta Astronautica}}, vol. 159, pp. 123--135, Jun. 2019.

\bibitem{Xia19beam}
S.~Xia, Q.~Jiang, C.~Zou, and G.~Li, ``{Beam coverage comparison of LEO
  satellite systems based on user diversification},'' \emph{{IEEE} Access},
  vol.~7, pp. 181\,656--181\,667, Dec. 2019.

\bibitem{R1.1908818}
{3GPP R1-1908818}, ``{Timing advance and RACH aspect for NTN},'' in \emph{{3GPP
  TSG RAN WG1 Meeting 98}}, Prague, Czech Republic, Aug. 2019.

\bibitem{R1.1910982}
{3GPP R1-1910982}, ``{On NTN synchronization, random access, and timing
  advance},'' in \emph{{3GPP TSG RAN WG1 Meeting 98bis}}, Chongqing, China,
  Oct. 2019.

\bibitem{R1.1910064}
{3GPP R1-1910064}, ``{Discussion on Doppler compensation, timing advance and
  RACH for NTN},'' in \emph{{3GPP TSG RAN WG1 Meeting 98bis}}, Reno, USA, Oct.
  2019.

\bibitem{3gpp.38.300}
{3GPP TS 38.300 V16.0.0}, ``{3rd Generation Partnership Project; Technical
  Specification Group Radio Access Network; NR; NR and NG-RAN Overall
  Description; Stage 2 (Release 16) },'' Tech. Rep., Dec. 2019.

\bibitem{Near}
W.~Wang, Y.~Tong, L.~Li, A.~Lu, L.~You, and X.~Q. Gao, ``{Near optimal timing
  and frequency offset estimation for 5G integrated LEO satellite communication
  system},'' \emph{{IEEE Access}}, vol.~7, pp. 3298--3310, Aug. 2019.

\bibitem{Morelli16robust}
M.~Morelli and M.~Moretti, ``{A robust maximum likelihood scheme for PSS
  detection and integer frequency offset recovery in LTE systems},''
  \emph{{IEEE} Trans. Wireless Commun.}, vol.~15, no.~2, pp. 1353--1363, Feb.
  2016.

\bibitem{Hsieh99low}
M.-H. Hsieh and C.-H. Wei, ``{A low-complexity frame synchronization and
  frequency offset compensation scheme for OFDM systems over fading
  channels},'' \emph{{IEEE} Trans. Veh. Technol.}, vol.~48, no.~5, pp.
  1596--1609, Mar. 1999.

\bibitem{Minn03robust}
H.~Minn, V.~K. Bhargava, and K.~B. Letaief, ``{A robust timing and frequency
  synchronization for OFDM systems},'' \emph{{IEEE} Trans. Wireless Commun.},
  vol.~2, no.~4, pp. 822--839, Jul. 2003.

\bibitem{3gpp.38.821}
{3GPP TR 38.821 V16.0.0}, ``{3rd Generation Partnership Project; Technical
  Specification Group Radio Access Network; NR; Solutions for NR to Support
  Non-terrestrial Networks (NTN) (Release 16) },'' Tech. Rep., Dec. 2019.

\bibitem{SatelliteOrbits}
O.~Montenbruck and E.~Gill, \emph{{Satellite Orbits-Models, Methods, and
  Applications}}.\hskip 1em plus 0.5em minus 0.4em\relax New York, NY, USA:
  Springer-Verlag, 2000.

\bibitem{pc82emitter}
P.~C. {Chestnut}, ``Emitter location accuracy using {TDOA} and differential
  doppler,'' \emph{{IEEE} Trans. Aerosp. Electron. Syst.}, vol. AES-18, no.~2,
  pp. 214--218, Mar. 1982.

\bibitem{Feng16app}
F.~{Shu}, S.~{Yang}, Y.~{Qin}, and J.~{Li}, ``Approximate analytic
  quadratic-optimization solution for {TDOA}-based passive multi-satellite
  localization with earth constraint,'' \emph{{IEEE} Access}, vol.~4, pp.
  9283--9292, Dec. 2016.

\bibitem{Feng18on}
F.~{Shu}, S.~{Yang}, J.~{Lu}, and J.~{Li}, ``On impact of earth constraint on
  {TDOA}-based localization performance in passive multisatellite localization
  systems,'' \emph{{IEEE} Syst. J.}, vol.~12, no.~4, pp. 3861--3864, Dec. 2018.

\bibitem{Jor04Num}
J.~Nocedal and S.~J. Wright, \emph{{Numerical Optimization}}, 2nd~ed.\hskip 1em
  plus 0.5em minus 0.4em\relax New York, NY, USA: Springer-Verlag, 2006.

\bibitem{nonli06Ba}
M.~S. Bazaraa, H.~D. Sherali, and C.~M. Shetty, \emph{{Nonlinear Programming:
  Theory and Algorithms}}, 3rd~ed.\hskip 1em plus 0.5em minus 0.4em\relax
  Hoboken, NJ, USA: John Wiley \& Sons, 2006.

\bibitem{nonli99Di}
D.~P.~Bertsekas, \emph{{Nonlinear Programming}}, 2nd~ed.\hskip 1em plus 0.5em
  minus 0.4em\relax Belmont, MA, USA: Athena Scientific Press, 1999.

\bibitem{funda93kay}
S.~Kay, \emph{{Fundamentals of Statistical Signal Processing, volume I:
  Estimation Theory}}.\hskip 1em plus 0.5em minus 0.4em\relax Englewood Cliffs,
  NJ, USA: Prentice Hall, 1993.

\bibitem{Asimplederivation}
T.~L. Marzetta, ``{A simple derivation of the constrained multiple parameter
  Cramer-Rao bound},'' \emph{{IEEE} Trans. Signal Process.}, vol.~41, no.~6,
  pp. 2247--2249, Jun. 1993.

\bibitem{zhao19multi}
B.~{Zhao}, G.~{Ren}, and H.~{Zhang}, ``Multisatellite cooperative random access
  scheme in low earth orbit satellite networks,'' \emph{{IEEE} Syst. J.},
  vol.~13, no.~3, pp. 2617--2628, Sep. 2019.

\bibitem{perf20zohair}
Z.~Abu-Shaban, G.~Seco-Granados, C.~R.~Benson, and H.~Wymeersch, ``Performance
  analysis for autonomous vehicle {5G}-assisted positioning in
  {GNSS}-challenged environments,'' in \emph{{Proc. 2020 IEEE/ION PLANS}},
  Portland, OR, USA, 2020, pp. 996--1003.

\bibitem{GDOP09sh}
I.~{Sharp}, K.~{Yu}, and Y.~J. {Guo}, ``{GDOP} analysis for positioning system
  design,'' \emph{{IEEE} Trans. Veh. Technol.}, vol.~58, no.~7, pp. 3371--3382,
  Sep. 2009.

\end{thebibliography}

\end{document}